\documentclass[twocolumn]{aastex631}
\newcommand{\Ha}{\hbox{{\rm H}\kern 0.1em$\alpha$}}
\newcommand{\Hb}{\hbox{{\rm H}\kern 0.1em$\beta$}}
\newcommand{\MgII}{\hbox{{\rm Mg}\kern 0.1em{\sc ii}}}
\newcommand{\CIV}{\hbox{{\rm C}\kern 0.1em{\sc iv}}}
\newcommand{\NeV}{\hbox{[{\rm Ne}\kern 0.1em{\sc v}]}}
\newcommand{\OII}{\hbox{[{\rm O}\kern 0.1em{\sc ii}]}}
\newcommand{\NeIII}{\hbox{[{\rm Ne}\kern 0.1em{\sc iii}]}}
\newcommand{\OIII}{\hbox{[{\rm O}\kern 0.1em{\sc iii}]}}
\newcommand{\NII}{\hbox{[{\rm N}\kern 0.1em{\sc ii}]}}
\newcommand{\SII}{\hbox{[{\rm S}\kern 0.1em{\sc ii}]}}

\newcommand{\lmass}{$\log\mathrm{M\!_\star/M}_{\odot}$}

\usepackage{color}
\definecolor{citeRGB}{rgb}{0,0.1,0.7}
\begin{document}

\title{Extremely red galaxies at $z=5-9$ with MIRI and NIRSpec: dusty galaxies or obscured AGNs?}

%\correspondingauthor{Guillermo Barro}
%\email{gbarro@pacific.edu}

\author[0000-0001-6813-875X]{Guillermo Barro}
\affiliation{University of the Pacific, Stockton, CA 90340 USA}
\author[0000-0003-4528-5639]{Pablo G. P\'erez-Gonz\'alez}
\affiliation{Centro de Astrobiolog\'{\i}a (CAB), CSIC-INTA, Ctra. de Ajalvir km 4, Torrej\'on de Ardoz, E-28850, Madrid, Spain}
\author[0000-0002-8360-3880]{Dale D. Kocevski}
\affiliation{Department of Physics and Astronomy, Colby College, Waterville, ME 04901, USA}
\author[0000-0001-8688-2443]{Elizabeth J.\ McGrath}
\affiliation{Department of Physics and Astronomy, Colby College, Waterville, ME 04901, USA}
\author[0000-0002-1410-0470]{Jonathan R. Trump}
\affiliation{Department of Physics, 196 Auditorium Road, Unit 3046, University of Connecticut, Storrs, CT 06269, USA}

\author[0000-0002-6386-7299]{Raymond C. Simons}
\affil{Department of Physics, 196A Auditorium Road, Unit 3046, University of Connecticut, Storrs, CT 06269, USA}

\author[0000-0002-6748-6821]{Rachel S. Somerville}
\affiliation{Center for Computational Astrophysics, Flatiron Institute, 162 5th Avenue, New York, NY, 10010, USA}

\author[0000-0003-3466-035X]{{L. Y. Aaron}  {Yung}}
\altaffiliation{NASA Postdoctoral Fellow}
\affiliation{Astrophysics Science Division, NASA Goddard Space Flight Center, 8800 Greenbelt Rd, Greenbelt, MD 20771, USA}

\author[0000-0002-7959-8783]{Pablo Arrabal Haro}
\affiliation{NSF's National Optical-Infrared Astronomy Research Laboratory, 950 N. Cherry Ave., Tucson, AZ 85719, USA}

 \author[0000-0002-9921-9218]{Michaela B. Bagley}
\affiliation{Department of Astronomy, The University of Texas at Austin, Austin, TX, USA}

\author[0000-0001-7151-009X]{Nikko J. Cleri}
\affiliation{Department of Physics and Astronomy, Texas A\&M University, College Station, TX, 77843-4242 USA}
\affiliation{George P.\ and Cynthia Woods Mitchell Institute for Fundamental Physics and Astronomy, Texas A\&M University, College Station, TX, 77843-4242 USA}

\author[0000-0001-6820-0015]{Luca Costantin}
\affiliation{Centro de Astrobiolog\'{\i}a (CAB/CSIC-INTA), Ctra. de Ajalvir km 4, Torrej\'on de Ardoz, E-28850, Madrid, Spain}

\author[0000-0001-8047-8351]{Kelcey Davis}
\affiliation{Department of Physics, 196 Auditorium Road, Unit 3046, University of Connecticut, Storrs, CT 06269, USA}

\author[0000-0001-5414-5131]{Mark Dickinson}
\affiliation{NSF's National Optical-Infrared Astronomy Research Laboratory, 950 N. Cherry Ave., Tucson, AZ 85719, USA}

\author[0000-0001-8519-1130]{Steve L. Finkelstein}
\affiliation{Department of Astronomy, The University of Texas at Austin, Austin, TX, USA}

\author[0000-0002-7831-8751]{Mauro Giavalisco}
\affiliation{University of Massachusetts Amherst, 710 North Pleasant Street, Amherst, MA 01003-9305, USA}

\author[0000-0002-4085-9165]{Carlos G{\'o}mez-Guijarro}
\affil{Universit{\'e} Paris-Saclay, Université Paris Cit{\'e}, CEA, CNRS, AIM, 91191, Gif-sur-Yvette, France}

\author[0000-0001-6145-5090]{Nimish P. Hathi}
\affiliation{Space Telescope Science Institute, Baltimore, MD, USA}

\author[0000-0002-3301-3321]{Michaela Hirschmann}
\affiliation{Institute of Physics, Laboratory of Galaxy Evolution, Ecole Polytechnique F\'ed\'erale de Lausanne (EPFL), Observatoire de Sauverny, 1290 Versoix, Switzerland}

\author[0000-0003-3596-8794]{Hollis B. Akins}
\affiliation{Department of Astronomy, The University of Texas at Austin, Austin, TX, USA}

\author[0000-0002-4884-6756]{Benne W. Holwerda}
\affil{Physics \& Astronomy Department, University of Louisville, 40292 KY, Louisville, USA}

\author[0000-0002-1416-8483]{Marc Huertas-Company}
\affil{Instituto de Astrof\'isica de Canarias, La Laguna, Tenerife, Spain}
\affil{Universidad de la Laguna, La Laguna, Tenerife, Spain}
\affil{Universit\'e Paris-Cit\'e, LERMA - Observatoire de Paris, PSL, Paris, France}

\author[0000-0003-1581-7825]{Ray A. Lucas}
\affiliation{Space Telescope Science Institute, 3700 San Martin Dr., Baltimore, MD 21218, USA}

\author[0000-0001-7503-8482]{Casey Papovich}
\affiliation{Department of Physics and Astronomy, Texas A\&M University, College
Station, TX, 77843-4242 USA}
\affiliation{George P.\ and Cynthia Woods Mitchell Institute for
 Fundamental Physics and Astronomy, Texas A\&M University, College
 Station, TX, 77843-4242 USA}

\author[0000-0001-7755-4755]{Lise-Marie Seill\'e}
\affiliation{Aix Marseille Univ, CNRS, CNES, LAM Marseille, France}

\author[0000-0002-8224-4505]{Sandro Tacchella}
\affiliation{Kavli Institute for Cosmology, University of Cambridge, Madingley Road, Cambridge, CB3 0HA, UK}
\affiliation{Cavendish Laboratory, University of Cambridge, 19 JJ Thomson Avenue, Cambridge, CB3 0HE, UK}

\author[0000-0003-3903-6935]{Stephen M.~Wilkins} %
\affiliation{Astronomy Centre, University of Sussex, Falmer, Brighton BN1 9QH, UK}
\affiliation{Institute of Space Sciences and Astronomy, University of Malta, Msida MSD 2080, Malta}

\author[0000-0002-6219-5558]{Alexander de la Vega}
\affiliation{Department of Physics and Astronomy, University of California, 900 University Ave, Riverside, CA 92521, USA}

\author[0000-0001-8835-7722]{Guang Yang}
\affiliation{Kapteyn Astronomical Institute, University of Groningen, P.O. Box 800, 9700 AV Groningen, The Netherlands}
\affiliation{SRON Netherlands Institute for Space Research, Postbus 800, 9700 AV Groningen, The Netherlands}

\author[0000-0002-7051-1100]{Jorge A. Zavala}
\affiliation{National Astronomical Observatory of Japan, 2-21-1 Osawa, Mitaka, Tokyo 181-8588, Japan}

\begin{abstract} 

We study a new population of extremely red objects (EROs) recently discovered by JWST based on their NIRCam colors F277W$-$F444W~$>1.5$ mag. We find 37 EROs in the CEERS field with F444W~$<28$ mag and photometric redshifts between $5<z<7$, with median $z=6.9^{+1.0}_{-1.6}$. Surprisingly, despite their red long-wavelength colors, these EROs have blue short-wavelength colors  (F150W$-$F200W$\sim$0~mag) indicative of bimodal SEDs with a red, steep slope in the rest-frame optical, and a blue, flat slope in the rest-frame UV. Moreover, all these EROs are unresolved, point-like sources in all NIRCam bands. We analyze the spectral energy distributions of 8 of them with MIRI and NIRSpec observations using stellar population models and AGN templates. We find that a dusty galaxy or an obscured AGN provide similarly good SED fits but different stellar properties: massive and dusty, \lmass$\sim$10 and A$_{\rm V}\gtrsim3$~mag, or low mass and obscuration, \lmass$\sim$7.5 and A$_{\rm V}\sim0$~mag, hosting an obscured QSO. SED modeling does not favor either scenario, but their unresolved sizes are more suggestive of an AGN. If any EROs are confirmed to have \lmass$\gtrsim10.5$, it would increase pre-JWST number densities at $z>7$ by up to a factor $\sim$60. Similarly, if they are OSOs with luminosities in the L$_{\rm bol}>10^{46-47}$ erg s$^{-1}$ range, their number would exceed that of bright blue QSOs by more than two orders of magnitude. Additional photometry at mid-IR wavelengths will reveal the true nature of the red continuum emission in these EROs and will place this puzzling population in the right context of galaxy evolution.

\end{abstract}

\section{Introduction}
\label{s:intro}

The extraordinary capabilities of the James Webb Space Telescope (JWST) provide the opportunity to completely transform our understanding of the high redshift Universe. The enhanced photometric sensitivity and spatial resolution at mid-infrared wavelengths relative to the Hubble Space Telescope (HST) or Spitzer have enabled, in the first few months of operations, a number of studies that have pushed the limits of the youngest and most distant galaxies detected in the epoch of reionization (e.g., \citealt{castellano22}; \citealt{naidu22b}; \citealt{adams23}; \citealt{whitler23}; \citealt{finkelstein23}; \citealt{pgp23a,pgp23b}) as well as expanded our identification of more massive galaxies up to $z\sim6$ and beyond (e.g., \citealt{tacchella22a}; \citealt{labbe23}; \citealt{nelson22}, \citealt{endsley22}). In the process, these papers have started to reveal the nature of the most massive galaxies that were previously undetected by HST (HST-dark) and detected only by Spitzer/IRAC, longer radio, and sub-mm wavelengths (\citealt{barrufet23}; \citealt{zavala23}; \citealt{pgp23a}; \citealt{rodighiero23}; \citealt{gomezguijarro23}) or not at all.

However, as we work our way towards a more complete census of the high-redshift Universe, there is a concern that some of these early estimates of the number density of galaxies or their (large) stellar masses could be in tension with model predictions (e.g. \citealt{mason23}; \citealt{ferrara22}; \citealt{boylan23}). A potential caveat for these photometric studies is that as we probe galaxies in the first 1 Gyr of the lifetime of the Universe we might find a large number of young, low-mass galaxies with extreme emission lines and potentially large equivalent widths (EW) of more than EW$=100 - 1000$\AA, as suggested by early studies of faint $z=5-7$ galaxies with Spitzer/IRAC (e.g., \citealt{egami05}; \citealt{eyles07}; \citealt{stark09}; \citealt{gonzalez12}; \citealt{labbe13}). Such large EWs can make the \Hb, \OIII, and \Ha\ line fluxes boost the broad and medium band photometry in the JWST/NIRCam filters up to F444W, making them appear very red. The impact on the colors can affect both the photometric redshifts (e.g., \citealt{arrabal23}) and the stellar population properties of these young, blue galaxies introducing a bias toward older ages, more dust obscuration, and significantly larger masses (\lmass$>10$). Recent JWST-based papers have reported that emission lines with large EW$\gtrsim$1000~\AA\ contaminating the NIRCam photometry are indeed a common occurrence (\citealt{endsley21,endsley22}; \citealt{matthee22}; \citealt{rinaldi23}) which may hamper the identification of true massive galaxies at $z>5$. Another potential concern with massive galaxy selections based on extremely red colors is the contamination by obscured active galactic nuclei (AGN). As shown also in IRAC-based studies, the red, power-law-like emission of an obscured AGN can also lead to very red optical to IR colors, which have been widely used to identify these galaxies in cosmological surveys (e.g., \citealt{alonsoherrero04}; \citealt{stern05}; \citealt{lacy07}; \citealt{donley08, donley12}). While the incidence of emission line or AGN contamination in color-selected samples at low-to-mid redshifts is only minor, the impact on JWST-based surveys is still unclear.  

A way forward to overcome the degeneracy in the origin of colors in red galaxies (high EW emission lines vs. stellar or AGN continuum) is to obtain photometry in multiple bands and extend the coverage to longer wavelengths. Clear detections at wavelengths that are not severely affected by strong emission lines would be a clear confirmation of continuum emission. Likewise, long wavelength detections probing the rest-frame near IR of the galaxies can help distinguish between the power-law AGN emission and the stellar 1.6~$\mu$m bump (\citealt{sawicki02}; \citealt{donley07}). Observations with JWST/MIRI at $\lambda>5\mu$m help break both of these degeneracies. Similarly, JWST/NIRSpec can provide precise redshifts for these galaxies and help calibrate the impact of the emission lines in photometric observations.

In this paper, we use the first and second epochs of data from the Cosmic Evolution Early Release Science Survey \citep{ceersers} to identify candidates for massive dusty galaxies at $z>5$ with very red colors in the long-wavelength NIRCam filters. Then, we focus on a subset of those galaxies with MIRI and NIRSpec observations to place better constraints on their redshifts and their emission at longer wavelengths, and we perform a detailed analysis of different spectral energy distribution (SED) modeling scenarios to determine the likelihood that they are blue high EW galaxies, dusty massive galaxies, or obscured AGN and the implications for the stellar masses and number densities of the sample in each case. 

The paper is structured as follows. In \S\ref{s:data}, we describe the data reduction of the multi-band NIRCam and MIRI imaging and the NIRSpec spectroscopy. We also describe the photometric measurement, catalog creation, and preliminary estimates of the photometric redshift and stellar properties for the whole CEERS region. In \S\ref{s:sample} and \S\ref{s:properties}, we perform the ERO color selection and we describe the colors, SEDs, photometric redshift and stellar masses of the sample selected that way. In \S\ref{s:stellarpop}, we perform a detailed SED modeling of a subset of 8 EROs observed with MIRI and NIRSpec using a variety of SED models aimed at testing the dusty galaxy vs. obscured AGN scenarios and their implications on the stellar population properties. In \S\ref{s:discussion}, we discuss the likelihood of the different modeling scenarios based on the general properties of the EROs as well as their best-fit SEDs. Lastly, We summarize our results and discuss future prospects in \S\ref{s:summary}.

Throughout this paper, we assume a cosmology with $H_0=70$~km~s$^{-1}$~Mpc$^{-1}$, $\Omega_M=0.3$, and $\Omega_{\Lambda}=0.7$. Quoted uncertainties are at the $1\sigma$\ (68\%) confidence level.
All magnitudes are in AB units \citep{oke83}.

\section{Data}
\label{s:data}

This paper is based on observations from the Cosmic Evolution Early Release Science Survey (CEERS), an early release science program \citep{ceersers} which covers approximately 100 arcmin$^{2}$ of the Extended Groth Strip (EGS) with imaging and spectroscopy using coordinated, overlapping parallel observations by multiple JWST instruments. Here we use the data acquired in June and December of 2022 which comprises 10 NIRCam pointings in 7 filters: 3 at short wavelengths (SW) F115W, F150W, F200W, and 4 at long wavelengths (LW) F277W, F356W, F410M, and F444W and 8 MIRI pointings in 7 filters, F560W, F770W, F1000W, F1280W, F1500W, F1800W, F2100W. Due to the nature of the CEERS parallel observations, some of the MIRI pointings are only observed either in the short (F560W, F770W) or long (F1000W to F2100W) wavelength filters and only 6 of them overlap with the NIRCam imaging. The names of these pointings in the APT observing file are 3, 6, 7 and 9, observed in F560W and F770W, and 5 and 8, observed at longer wavelengths only. In addition to NIRCam imaging, pointing 3, 6 and 7 overlap with the NIRCam WFSS grism observations and two of the NIRSpec pointings named 9 and 10 in the APT file.

The NIRCam and MIRI data were calibrated using version 1.7.2 of the JWST Calibration Pipeline, reference files in pmap version 0214 (which includes detector-to-detector matched, improved absolute photometric calibration), with some additional modifications described in more detail in \citet{finkelstein23} and \citet{bagley23} for NIRCam and \citet{papovich23} and Yang et al. (in prep) for MIRI.  The reduced images are registered to the same World Coordinate System reference frame (based on Gaia DR1.2; \citealt{gaia16}) and coadded into single mosaics with pixel scales of 0.$\arcsec$03/pixel and 0.$\arcsec$09/pixel for NIRCam and MIRI, respectively.

The CEERS NIRSpec observations (Arrabal Haro in prep.) were processed using version 1.8.5 of the JWST Science Calibration Pipeline, with the Calibration Reference Data System (CRDS) mapping 1027 following similar procedures as in \citet{fujimoto23} and \citet{kocevski23}. Briefly, we correct for the detector 1/f noise, subtract the dark current and bias, and generate count-rate maps starting from the uncalibrated images. We apply a few additional custom steps to improve the treatment of the cosmic ray ``snowballs". The resulting maps are processed with stage two of the pipeline to generate reduced two-dimensional (2D) spectra with a rectified trace and flat slope. Custom extraction apertures are determined visually by inspecting the images for high signal-to-noise ratio continuum or emission lines. Lastly, we extract the 1D spectra boxcar apertures centered on the visually identified trace.

\subsection{Source extraction and photometry}

The source extraction and multi-band photometric measurement were performed following the same methods as for the first epoch data described in detail in \citet{finkelstein23}. Briefly, photometry was computed on PSF-matched images using SExtractor \citep{sex} v2.25.0 in two-image mode, with an inverse-variance weighted combination of the PSF-matched F277W and F356W images as the detection image. Photometry was measured in all seven of the NIRCam bands observed by CEERS, as well as the F606W, F814W, F105W, F125W, F140W, and F160W HST bands using data obtained by the CANDELS and 3D-HST surveys (\citealt{candelsgro}; \citealt{candelskoe}; \citealt{brammer11}).

\subsection{Circular aperture photometry}
\label{s:aperturephot}

We recompute the photometry of the subsample of objects studied in \S~\ref{s:stellarpop} using smaller circular apertures to improve the precision in the photometric errors and to avoid  potential photometric contamination by nearby sources or background subtraction problems. Given that the nature of our galaxies is very homogeneous, all sources analyzed in this paper are barely resolved or unresolved (see \S\ref{s:properties}), photometric apertures with 0.4\arcsec\, diameter were the most adequate to obtain the most precise and reliable SEDs. Photometry was measured in original and PSF-matched images, and after applying aperture corrections for point-like sources for the former, we arrived at consistent colors within at least half the value of the photometric corrections.

\subsection{Photometric redshifts and stellar population properties}
\label{s:photoz}

We estimate photometric redshifts for the whole parent catalog by fitting the multi-band SEDs using the code \texttt{EAZYpy} \citep{eazy}. The code fits non-negative linear combinations of templates to the observed data to derive probability distribution functions (PDFs). Here we use the default template set ``tweak fsps QSF 12 v3" which consists of a set of 12 templates derived from the stellar population synthesis code FSPS \citep{conroy10}. As a result, in addition to the photometric redshift the code also provides an estimate of the stellar mass as well as the dust attenuation. In addition, we also estimate stellar population properties by fitting the optical and NIR SEDs using FAST \citep{kriek09}, assuming \citet{bc03} stellar population synthesis models, following a \citet{chabrier} IMF, a delayed exponential star formation history, and the \citet{calzetti} dust law with attenuation $0<$~A$_{\rm V}<4$~mag.

\begin{figure*}%[htp!]%[ht!]
\centering
\includegraphics[width=18.cm,angle=0]{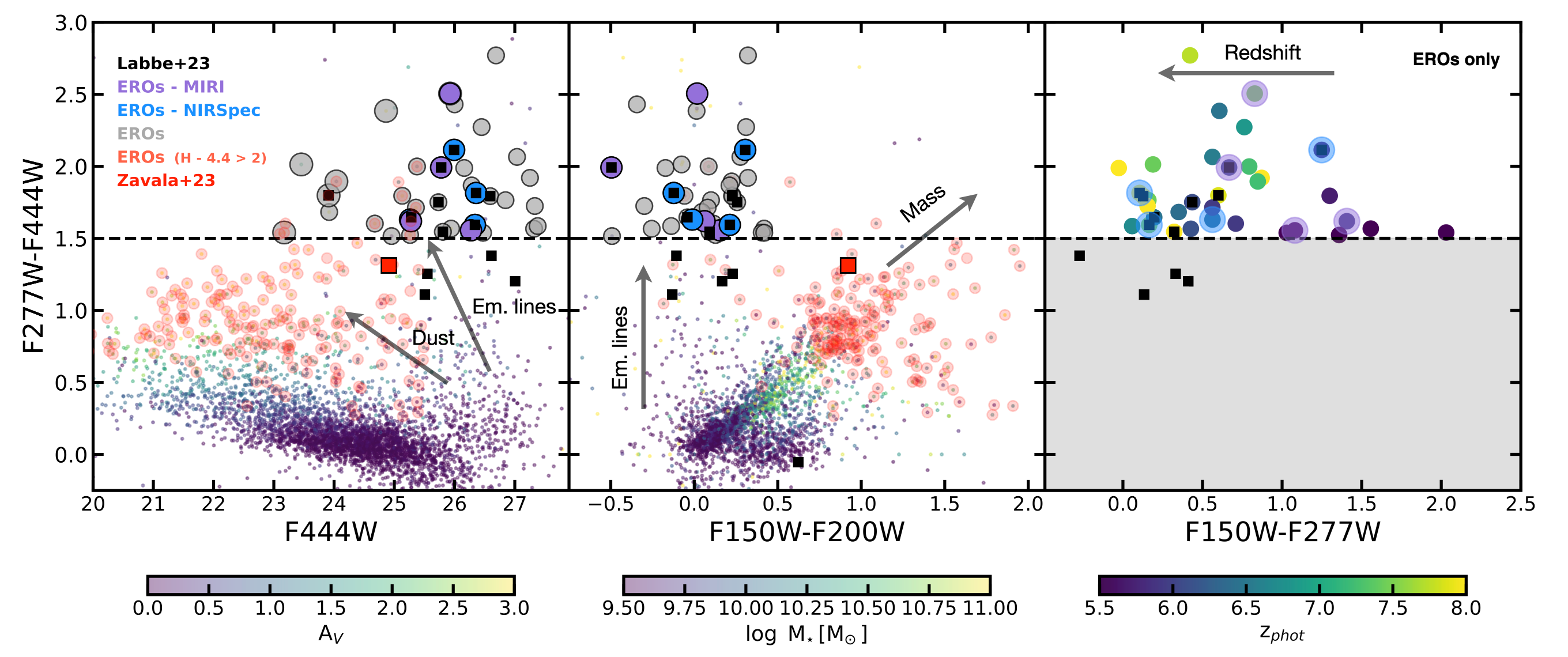}
\caption{Color-magnitude and color-color diagrams showing the selection threshold for
F277W EROs (circles; F277W$-$F444W$>$1.5~mag), relative to the bulk of the CEERS galaxy catalog, color-coded by stellar mass and A$_{\rm V}$, and a subset of F150W EROs (F150W$-$F444W$>$2~mag). The blue and purple markers indicate the EROs observed with MIRI and NIRSpec. The black squares show the EROs from \citet{labbe23}. The left and central panels show the general trends toward redder colors with increasing mass and dust attenuation (arrows), which suggest that F277W EROs are massive and dusty galaxies. However, the central panel reveals that F277W EROs have surprisingly blue colors at shorter wavelengths, F150W$-$F200W$\sim0$ mag, very different from those of F150W EROs and massive dusty galaxies in general. The red square shows a  massive, dusty, sub-mm galaxy at $z=5.1$ from \citet{zavala23} which is also red in all bands. This implies that F277W EROs have bimodal SEDs with blue SW colors and red LW colors. The right panel shows the correlation between photometric redshift and F150W$-$F277W color for the F277W EROs. As the F277W filter shifts from the steep, rest-frame optical range to the flat rest-UV range with increasing redshift the color declines to F150W$-$F277W$\sim$0~mag.}
\label{fig:sample}
\end{figure*}

\section{Sample Selection}
\label{s:sample}
\subsection{ERO color criterion}
\label{s:eroselect}

We identify extremely red galaxies at high redshift using a single color cut F277W$-$F444W~$>1.5$~mag.  This method is similar to the traditional ERO (R$-$K, e.g., \citealt{mccarthy04}) or IERO (K$-$[4.5], e.g., \citealt{caputi13}; \citealt{stefanon13}; \citealt{wang12}) selection which uses red optical to NIR colors to find massive, dusty, or quiescent galaxies with strong Balmer or 4000\AA~breaks at $z\gtrsim3$. With the arrival of JWST, this technique has been extended to fainter magnitudes and higher redshifts by using filters at longer wavelength, for example, F150W$-$F444W in \citet{barrufet23}, or F150W$-$F356W in \citet{pgp23a}. Recently, \citet{labbe23} used a threshold of F277W$-$F444W~$>1$~mag to identify candidates to massive galaxies at $z>7$. Here we use a slightly redder color and we drop the additional color constraints to lower the selection redshift to $z\gtrsim5$. A redder color threshold can also reduce the contamination by galaxies with high EW emission lines ($>1000$\AA) boosting the NIR fluxes of blue galaxies with a relatively shallow stellar continuum. For example, \citet{endsley22} find red colors, F277W$-$F444W$\lesssim1$ and F277W$-$F356W $\lesssim1$, in a sample of low mass, Lyman-break galaxy candidates at z$=6.5-8$ which were largely driven by high EW \OIII/\Hb\ lines boosting the flux in F444W. Such strong lines have also been spectroscopically confirmed by recent NIRCam/WFSS surveys at slightly lower redshifts of z$>$5.3  \citep{matthee22}. 

The left panel of Figure~\ref{fig:sample} illustrates the sample selection in a color-magnitude diagram compared to the overall distribution of galaxies in the CEERS catalog, color-coded by different properties, and a sub-sample of F150W EROS (F150W$-$F444W$>$2; F444W$<$25 mag; red circles). The 13 galaxies from  \citet{labbe23} are shown with black squares. All of them except the 4 with color F277W$-$F444W~$<$1.5~mag are included in our sample. The color code in the CEERS sample highlights the trend of increasing NIR colors with extinction (and similarly with stellar mass and redshift in the other panels). As discussed above, galaxies redder than the color threshold (dashed line) are candidates for massive galaxies with red, dusty, or quiescent SEDs and possibly some galaxies with high EWs emission lines. Interestingly, there are some differences between the sample of F150W EROs and F277W EROs. First, F277W EROs are fainter, with a median magnitude of F444W $=$ 25.9$^{+0.8}_{-1.1}$~mag, whereas F150W EROs span a broader range in magnitude starting at F444W$\gtrsim$20~mag which is consistent with the notion that by selecting in a redder band, F277W EROs lean more toward the higher redshift tail of the massive galaxy selection. Second, F150W EROs are typically selected within a brighter limiting magnitude to restrict the number of galaxies in the lower mass end of the selection criteria \lmass$\sim$10 (e.g., F444W$\lesssim25-26$; \citealt{alcalde19}; \citealt{gomezguijarro23}). However, using a fainter limiting magnitude increases the overlap between the two ERO samples, as shown for example in \citet{pgp23a}. Nevertheless, we find that, even within a similar magnitude range, the F150W selection misses some F277W EROs because they have bluer colors in F150W$-$F444W$=$2.2$^{+0.7}_{-0.5}$ mag. 

The reason for this key difference is highlighted in the central panel of Figure~\ref{fig:sample} which shows that all the F277W EROs are surprisingly blue in F150W$-$F200W$\sim$0~mag, which probes the rest-frame UV at z$>$5. Consequently, these EROs populate a very different region of the color-color diagram far from the loci of the F150W EROs, and all other massive galaxies which typically have red colors, F150W$-$F200W$=$0.5 to 1.5 mag. This means that, unlike the majority of other massive galaxies, which are red across their whole SEDs, the F277W EROs are blue in the rest-UV and red in the rest-optical. Such peculiar colors indicate that these EROs have bimodal blue-red SEDs (L-shaped or V-shaped in f$_{\lambda}$), as noted by \citet{labbe23}. The right panel shows that the goal of the second color threshold (F150W$-$270W$<0.7$~mag) in the selection method of \citet{labbe23} is to remove galaxies at $z<7$ from the sample. The F150W$-$F270W color acts as a pseudo redshift because the F277W filter shifts from the steep optical side of the SED to the flat UV with increasing redshift. Consequently, the color quickly declines toward F150W$-$F270W$\sim$0 for galaxies at $z\gtrsim7$. For the same reason, the primary selection in F277W$-$F444W might start missing galaxies of this type at $z\gtrsim9$ when the F444W filter starts to shift out the steep rest-optical range. Lastly, we note that the selection in F277W$-$F444W~$>1.5$~mag is surprisingly clean as it only identifies these peculiar EROs with bimodal, blue-red SEDs with no contamination from typical EROs (i.e., red across their whole SEDs). 

We identify 37 EROs with the color criterion described above. We visually inspect all the candidates and we remove some unreliable detections (e.g., hot pixels or fake objects extracted near bright stars of diffraction spikes). Their average magnitudes in F444W, F536W, F277W, and F150W are 25.9$^{+0.8}_{-1.1}$, 26.8$^{+0.9}_{-1.2}$, 27.6$^{+0.9}_{-1.4}$, and 28.2$^{+1.0}_{-1.3}$ mag, respectively, which is consistent with the color selection criterion. Their very faint magnitudes in F150W imply that these objects are all HST/WFC3 drop-outs at the depth of the CANDELS data in the CEERS region.

\subsection{MIRI detection and NIRSpec spectroscopy of the EROs}
\label{s:miri_nirspec_det}

We search for counterparts of the 37 EROs in the CEERS MIRI and NIRSpec observations. Unfortunately, the MIRI coverage of the CEERS/NIRCam mosaic is quite limited (less than $\sim$8\% of the area) and none of the pointings have simultaneous observations in the short and long wavelengths bands. Overall, only 4 of the MIRI pointings in F560W and F770W, and 2 of the pointings observed in F1000W and onward overlap with NIRCam coverage. Surprisingly, we find clear detections for all 4 of the 37 EROs which lie within the MIRI observed area. Three of them are detected in F560W and F770W with an average magnitude of 25.4$^{+0.3}_{-0.1}$ mag and 25.3$^{+0.2}_{-0.1}$ mag respectively and 1 of them is weakly but clearly detected in F1000W at 24.6~mag. While it is difficult to extrapolate from such a small sample, the high recovery fraction of observed objects, as well as the very red, power-law-like slope of the SED in the LW NIRcam bands, suggests that follow-up observations of similarly selected EROs in other fields with a more dense MIRI coverage is likely to yield a significant number of detections. Given the median magnitude of these objects F444W=25.9$^{+0.8}_{-1.1}$ mag we would expect detections in F560W and F777W in the 25 to 26~mag range which is clearly within the 5$\sigma$ limit for surveys similar to CEERS, like PRIMER or COSMOS-Web. Note also that above $z>7$, the \Ha\ emission line shifts into the F560W filter (see Figure~\ref{fig:stack_lines}) which might further enhance the flux and facilitate the detection.

In addition to the MIRI detections, 4 other EROs have been observed as part of the CEERS NIRSpec survey. All of them have clear emission lines that provide a robust estimate of their redshifts. Two of them have already been presented and discussed in \citet{fujimoto23} and \citet{kocevski23}, nircam3-2232 and nircam3-3210, respectively. The two galaxies at $z>7$ exhibit only \Hb\ and \OIII\ detections, while the other two a z$\lesssim$6 show \Ha\ as well. The galaxy discussed in \citet{kocevski23}, at $z=5.62$, is the only one that has a continuum detection and exhibits a broad-line \Ha\ emission which confirms that it is an AGN. All galaxies have relatively low \OIII/\Hb\ ratios, however, as noted \citet{kocevski23}, the narrow emission-line ratios are very similar to those of star-forming galaxies observed at similar redshifts which means that the line-ratio AGN diagnostic might not be particularly effective at $z\gtrsim5$.

\begin{figure*}
\centering
\includegraphics[width=18cm,angle=0]{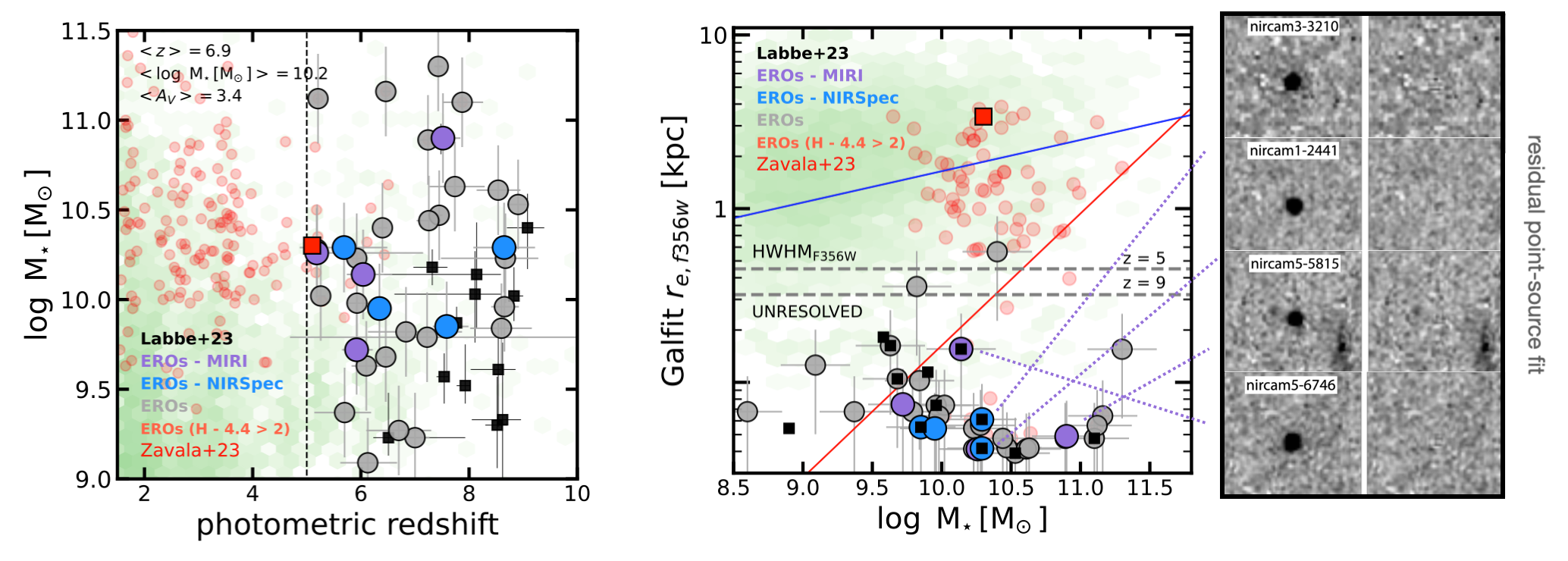}
\caption{{\it Left}: Photometric redshift vs. stellar mass diagram for the F277W EROs (circles), the CEERS galaxy sample (green density map), and the F150W EROs (red). The F277W EROs are relatively massive, \lmass$\sim$10, and dusty, A$_{V}\sim$3~mag and they span a redshift range from $5<z<9$. Overall, F277W EROs are among the most massive galaxies at their redshift, but less massive than the F150W EROs at lower redshift, following the expected decline in the number of very massive galaxies with redshift. However, a few of them are much more massive (\lmass$\gtrsim$10.5)m suggesting that there might be limitations in the  
fitting of their bimodal SEDs or perhaps that the continuum is not stellar, but AGN-dominated (see discussion in \S~\ref{s:stellarpop}). {\it Right}: Stellar mass vs. F356W effective radius for the same galaxies. The blue and red lines show the mass-size relations for star-forming and quiescent galaxies from \citet{vdw14}. The dashed lines indicate the approximate resolution limit from the HWHM of the PSF in F356W (FHWM=0.15\arcsec) at $z=5$ and $z=9$. Remarkably, all the F277W EROs appear to be unresolved point-like sources in contrast with the typical spread of F150W EROs and other massive galaxies. We find similar results in the other NIRCam bands suggesting that the EROs are unresolved at all wavelengths. The panels on the right show the best fit to a PSF in F444W for the 4 galaxies with MIRI detections showing negligible residuals.}
\label{fig:photoz_size}
\end{figure*}

\section{Properties of the  EROs}
\label{s:properties}

\subsection{Photometric redshifts, and stellar masses}

The left panel of Figure~\ref{fig:photoz_size} shows the overall distribution of the F277W EROs in photometric redshift and stellar mass compared to the bulk of the CEERS sample (green density map) and the sample of F150W EROs from Figure~\ref{fig:sample} (red). Overall, the F277W EROs are relatively massive and dusty with median values of \lmass $=$ 10.2$^{+0.5}_{-0.4}$ and A$_{\rm V}$=3.0$^{+1.3}_{-1.1}$~mag, similar to those reported in \citet{labbe23} for the $z>7$ population. The redshift distribution ranges between $5<z\lesssim9$ with a median of $z=6.9^{+1.0}_{-1.7}$. This indicates that nearly half the sample is at 
redshifts $5<z<7$, as suggested by \citet{pgp23a}. We remove a single object at $z\lesssim5$ for homogeneity, but the color selection is, overall, very effective at identifying galaxies at $z>5$. As expected from the color selection, the EROs tend to be among the most massive galaxies at their redshift  (i.e., relative to the green map). Compared to the other F150W EROs at lower redshift, the F277W EROs tend to follow the expected decline in the number of very massive galaxies, \lmass$\gtrsim$10.5, with redshift. However, we find a handful of galaxies with large masses, \lmass$\gtrsim$10.5, even at $z>7$ which, if confirmed, would be hard to reconcile with the observed stellar mass functions as well as models of galaxy evolution (e.g., see discussion in \citealt{boylan23}). Not surprisingly, these galaxies are also among the brightest in F444W by nearly 1 or 2 magnitudes relative to the median of the sample. The reliability of the stellar mass estimates is indeed one of the fundamental questions about these EROs with unusual SEDs. The values discussed in this section are computed with \texttt{FAST} based on typical modeling assumptions (see \S\ref{s:photoz}), which work well for most galaxies at low to mid redshifts. However, this method might have limitations for these EROs (e.g., because of strong emission lines or extreme obscurations). In \S~\ref{s:stellarpop} we analyze in detail the impact of using different codes and modeling assumptions on the inferred stellar masses. 

\subsection{Sizes and morphologies}

The right panel of Figure~\ref{fig:photoz_size} shows the distribution of the F277W EROs in the stellar mass vs. size diagram compared to F150W EROs at $z>3$ and the overall distribution of galaxies in the CANDELS F160W catalog in the overlapping area with CEERS (green density map). The CANDELS measurements are derived from \citealt{stefanon15} and \citealt{vdw14}. Sizes are represented by the effective radius, r$_{\rm e}$, of the S\'ersic \citep{1968ApJ...151..393S} profile fit performed with  GALFIT v3.0.5 \citep{galfit} in the F356W band. The code was run on the background-subtracted images with sizes 2.5 times the Kron radius. The ERR array, which includes background sky, Poisson, and read noise, was used as the input noise map. Empirical PSFs were constructed using stars in all CEERS pointings. All galaxies in the image cutout within 3 magnitudes of the primary source were fit simultaneously. All other sources were masked out during the fitting. The fitting parameters were allowed to vary within the following reasonable bounds: S\'ersic index $0.2\le n \le 8.0$, effective radius $0.3 \le r_e \le 400\rm{~pixels}$, axis ratio $0.01 \le q \le 1$, magnitude $\pm3$ mag from the initial value, and position $\pm3$ pixels from the initial value. 

Overall, we find that while the F150W EROs tend to overlap with the bulk of the galaxy sample, scattered in between the expected mass-size relations for star-forming and quiescent galaxies (blue and red lines from \citet{vdw14} at $z=3$), all the F277W EROs are extremely small, systematically under the resolution limits regardless of their stellar masses. The best-fit GALFIT r$_{\rm e}\sim$0.009\arcsec (0.3 px) returns in most cases the absolute lower limit set for the fitting suggesting that the galaxies are not resolved. The dashed lines indicate the approximate minimum sizes measurable as the HWHM of the PSF (0.07\arcsec) at $z=5$ and $z=9$, roughly $r_{\rm e}\sim$0.3$-$0.4~kpc. We further explore the size measurements of the EROs in F200W, F277W, and F444W obtaining similar results which suggest that they are unresolved in all the observed wavelengths. Note that the EROs are typically very faint ($\sim$27-28 mag) in all the SW NIRcam bands and, in most cases, they have only a handful of bright pixels for the fitting. Lastly, we also fit the profiles of the 8 EROs with MIRI and NIRSpec detections using point-like PSFs and we find excellent agreement with negligible residuals (right panel of Figure~\ref{fig:photoz_size}) indicating that they are indeed unresolved.

\subsection{Overall SEDs and possible modeling scenarios}

\begin{figure*}%[htp!]%[ht!]
\includegraphics[width=17.6cm,angle=0]{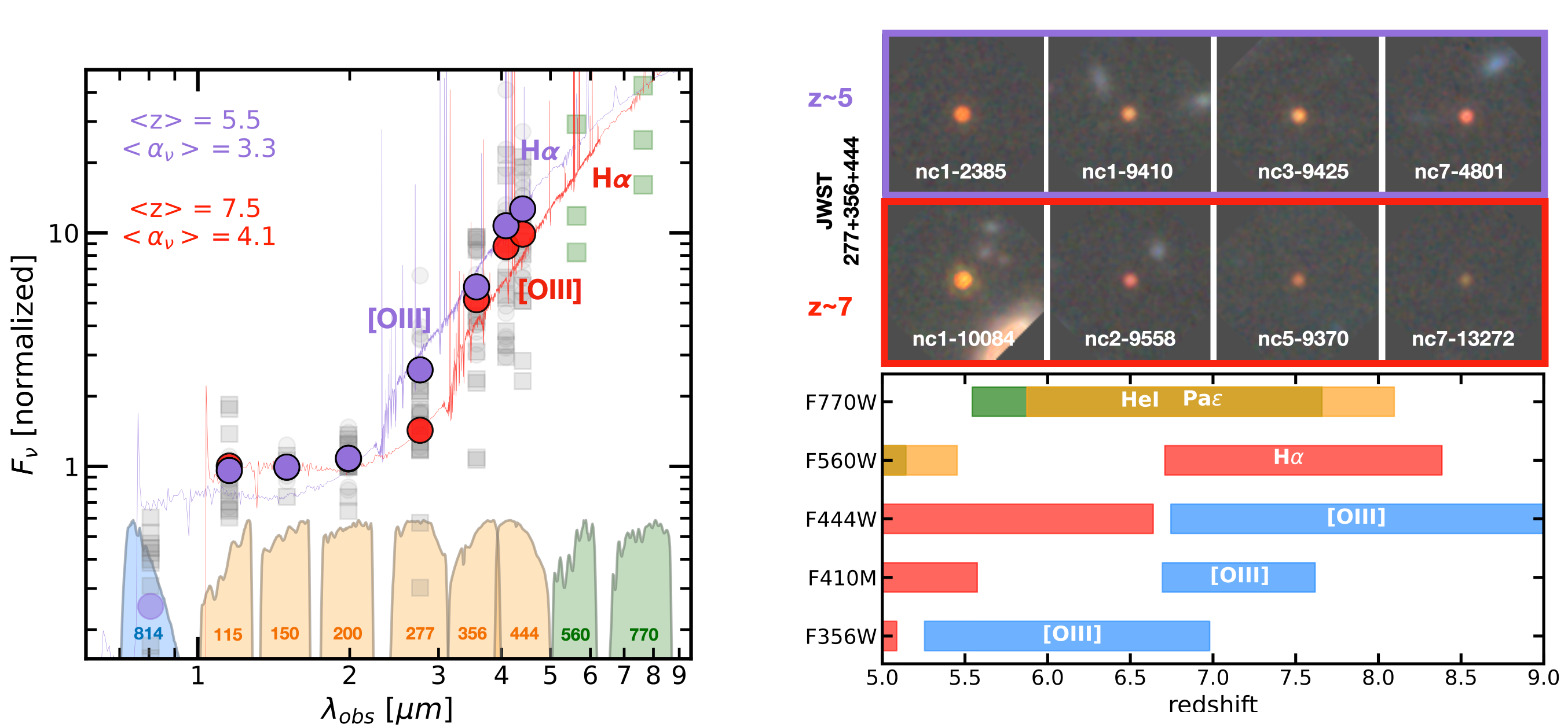}
\caption{{ \it Left:} Stacked spectral energy distribution of the 37 EROs (grey squares) divided into 2 groups below and above z=7, shown in purple and red, respectively. The MIRI photometry is shown in green. All galaxies exhibit a characteristic bimodal SED. Representative best-fit SEDs with \texttt{EAZYPy} at z$=$5.5 and z$=$7.5 (purple and red solid lines) show that this peculiar SED shape is typically reproduced by a composite SED with a blue, flat continuum in the rest-frame UV and red, steep continuum in the optical. Indeed, the best-fit power-law to the fluxes redward of F277W is quite large ($\alpha_{\nu}\sim3-4$) indicative of a heavily reddened continuum. The stacked SEDs also highlight the difference in F277W as the bands shift from the steep to the flat slope with increasing redshift. {\it Right-top:} 2.5\arcsec$\times$2.5\arcsec cutouts of EROs in the two redshift bins showing their similar compact and featureless visual appearances. {\it Right-bottom:} List of some of the strongest emission lines that can potentially cause emission-line-driven excesses in the NIRCAM and MIRI photometry at different redshifts. The locations of the strongest \Ha\ and \OIII\ lines are also indicated in the left panel.}
\label{fig:stack_lines}
\end{figure*}

The right panel of Figure~\ref{fig:stack_lines} shows the stacked SED of all the EROs normalized to the median of the relatively flat rest-UV continuum traced by F115W, F150W, and F200W divided into two groups at redshifts below and above $z=7$, purple and red markers respectively. Both groups exhibit the distinctive, bimodal SED discussed in \S~\ref{s:sample}, which consists of extremely red colors at  $\lambda>2$~$\mu$m, with a relatively constant power-law slope $\sim3.5\pm0.5$~$\mu$Jy/$\mu$m, and a flat SED at shorter wavelengths. The red, power-law-like emission is typically associated with large amounts of dust attenuation. However, as discussed in \S~\ref{s:eroselect}, it is also possible that the flux in some of the LW filters is partially boosted by strong emission lines, making the colors redder than the underlying stellar continuum. The right panel of Figure~\ref{fig:stack_lines} highlights the location of some of the strongest lines that can boost the emission in different filters as a function of redshift. At $5<z<7$, the H$\alpha$ and \OIII~lines can contaminate the F444W and F356W filters while F277W probes the continuum redward of the 4000\AA\ break. At $z>7$, the same lines shift into F444W and MIRI/F560W while F356W probes the red continuum. The average, stacked fluxes in F277W and F356W for the low and high redshift groups are both clearly above the flat continuum in the rest-UV suggesting that there is at least some continuum emission redward of 4000\AA . Furthermore, it would be difficult to reproduce a constant power-law slope spanning both the NIRCam and MIRI bands with relatively normal, low EW$\sim$100\AA\ emission lines since typically at least one, but probably several bands, should not be affected by the most prominent emission lines.
 
Nonetheless, the very pronounced change in the slope from the blue to the red spectral region is also difficult to model in terms of a single stellar continuum. Indeed, the best-fit templates from \texttt{EAZY-py} at $z=5.5$ and $z=7.5$ shown in the left panel of Figure~\ref{fig:stack_lines} are often composites of two templates with very different stellar ages, masses, and dust attenuations: on the one hand, a young, low-mass, low-attenuation galaxy (i.e., a typical LBG) and, on the other, a more massive and dusty galaxy. As a consequence, the inferred stellar mass and extinction of the composite is usually quite large, because it is dominated by the larger mass-to-light ratio of the older galaxy.

Recently, other works (e.g., \citealt{endsley22}; \citealt{furtak23}) discussed the possibility that the SEDs of some of these EROs could be explained partially, or completely, by very strong, AGN-driven emission lines. The presence of high EW$>1000$\AA\ emission lines can boost the flux in all the filters since these are not restricted to just the brightest emission lines due to star-formation. Similarly, the peculiar SEDs can also be explained in terms of continuum emission from an AGN which can outshine the galaxy host in different spectral regions. This possibility was recently explored in \citet{kocevski23} for one of the EROs at $z=5.62$ with NIRSpec observations, which is also included in our sample (nircam3-3210). This galaxy was also discussed in \citet{labbe23} but the estimated photo-z was much higher $z\sim8$. This highlights again the potential pitfalls in the SED modeling of these galaxies. \citet{kocevski23} proposed some AGN-dominated scenarios where the SED could be explained by: 1) a heavily obscured QSO dominating the LW fluxes and a small percentage of scattered light from the broad-line component causing the blue, SW emission (e.g., as in the \citealt{polletta06} torus template), 2) a heavily obscured QSO dominating the LW fluxes plus a blue, low-mass galaxy host, which dominates the SW fluxes, or 3) a blue, type 1 QSO, dominating the SW fluxes, in a dusty starburst galaxy which dominates the LW emission. The latter is also similar to the red QSO scenario in \citet{fujimoto22}.

Crucially, many of these different scenarios can be confirmed or ruled out with additional observations such as the NIRSpec spectroscopy in \citet{kocevski23} or with additional photometry at longer wavelengths from JWST/MIRI. For example, \citet{papovich23} and \citet{rinaldi23}, have recently shown that many of the blue, low-mass LBGs at $z>7$ with emission line-driven excesses in F444W have clear detections in MIRI at F560W and F777W that can trace the continuum in a spectral region without prominent emission lines. For these EROs, MIRI detections in the rest-frame optical continuum can distinguish between scenarios where the red optical colors are primarily driven by high EW emission line vs. any kind of continuum-dominated emission by a red, dusty galaxy or a QSO. In the \S~\ref{s:stellarpop}, we study the likelihood and implications of the different scenarios outlined above from a detailed analysis of the SED modeling of the 4 galaxies with additional photometric constraints from MIRI and the 4 galaxies with spectroscopic redshifts from NIRSpec. In \S~\ref{s:discussion} we use those results to inform the discussion on what would be the most likely scenario for the whole population of EROs.

\section{SED modeling of the MIRI and NIRSpec detected EROs}
\label{s:stellarpop}

\begin{figure}%[htp!]%[ht!]
\includegraphics[width=8.2cm,angle=0]{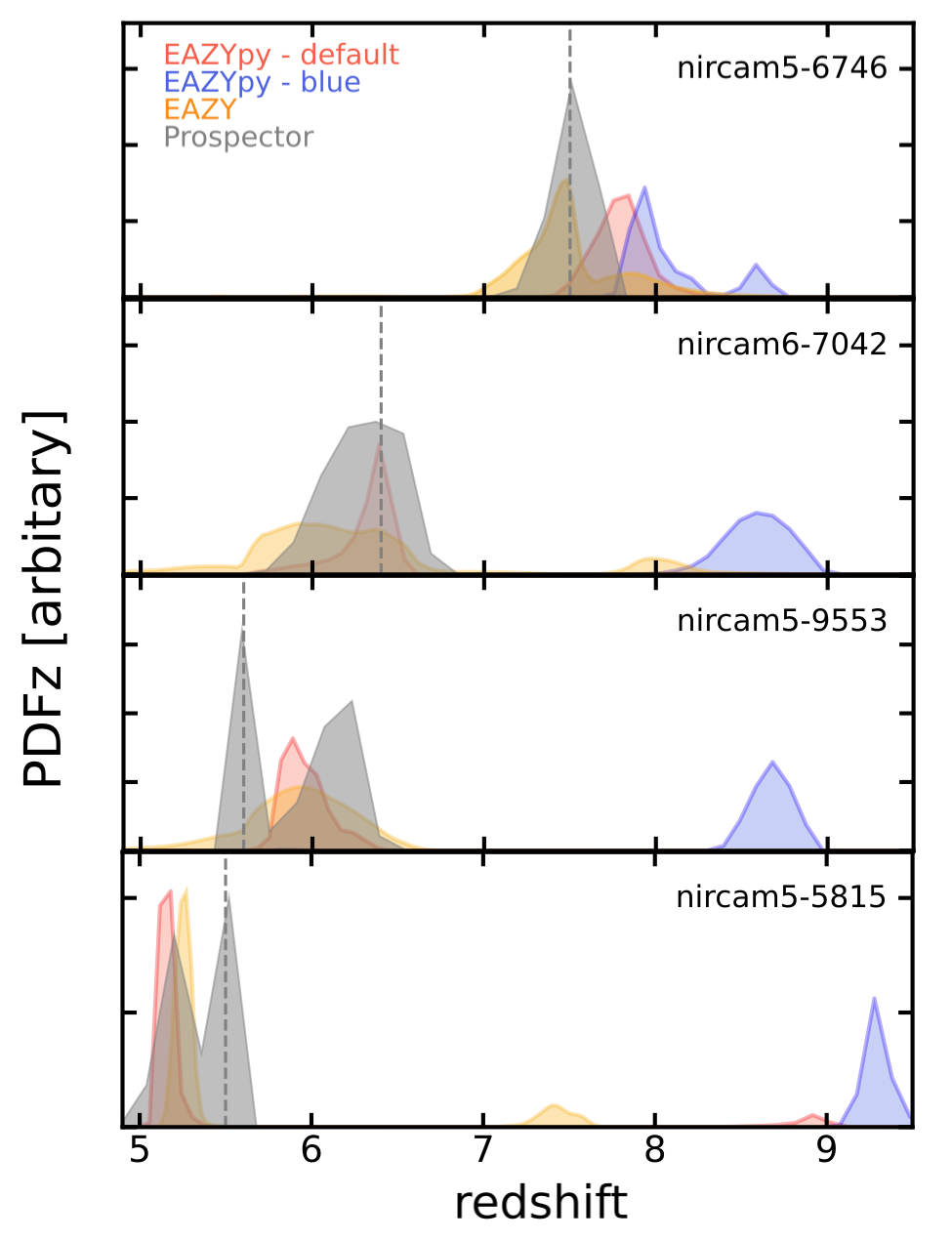}
\caption{Photometric redshift distributions (PDFz) for the 4 MIRI detected EROs computed using \texttt{EAZY}, \texttt{EAZYpy}, and \texttt{Prospector}. The PDFz's derived with the default and blue version of the \texttt{EAZYpy} templates agree well with one another and with the \texttt{Prospector} estimates for all the galaxies. For the 3 galaxies at $z<7$, the PDFz's based on the templates with very high EW lines (blue) suggest a secondary peak at higher redshift that is not supported by the detections in F150W. The key difference between the low and high-z peaks is typically an emission line-driven excess in F444W which could be attributed to \Ha\ or \OIII\ respectively (see also Figure~\ref{fig:stack_lines}).}
\label{fig:miri_photoz}
\end{figure}

\subsection{Modeling codes}
\label{s:codes}

In this section, we perform a more detailed SED modeling of the 8 EROs with MIRI and NIRSpec observations using the SEDs derived from the circular aperture photometry described in \S~\ref{s:aperturephot}
and a variety of codes aimed at exploring the likelihood of the different dusty galaxy vs. obscured AGN scenarios outlined in the previous section. A detailed description of the modeling assumptions adopted for each code is provided in Appendix 1. Briefly, we use \texttt{EAZYpy} \citep{eazy}, \texttt{Synthesizer} \citep{pg08}, \texttt{Prospector} \citep{johnson21}, and a custom code to perform a hybrid fit of the stellar population models from \texttt{Prospector} with the AGN templates of \citealt{polletta06}. The \texttt{EAZYpy} fits are based on the same default template set used in \S\ref{s:photoz}. The \texttt{Synthesizer} run uses parametric SFHs, following a delayed-$\tau$ function characterized with the \citet{bc03} stellar population models, a \citet{calzetti} attenuation law and nebular emission following \citet{cloudy}. With \texttt{Prospector}, we use 3 different options: 1) a fiducial model with a parametric delayed-$\tau$ SFHs and \citet{calzetti} attenuation law; 2) a non-parametric SFH based on the continuity priors of \texttt{Prospector-$\alpha$} (e.g., \citealt{leja19} or \citealt{tacchella22a}) but with a maximum age of 100~Myr and using a \citet{calzetti} attenuation law; 3) a similar non-parametric SFH with a more complex dust attenuation model based on \citet{cf00} and \citet{kriek13}. All 3 options are based on the FSPS models \citep{conroy09} and include nebular emission from young stars. They also have a number of other modeling assumptions in common (gas and stellar metalicity, ionization parameter, etc.) described in the appendix. The first 2 options are aimed at exploring the impact of using parametric/non-parametric SFHs and different stellar population models with respect to \texttt{Synthesizer}, while the third focuses on the impact of the dust attenuation law.
The last SED model is a hybrid of a galaxy and a dust-obscured QSO. Here we assume that the emission in the LW NIRCam and MIRI bands is largely dominated by an obscured QSO modeled after the QSO2 from \citet{polletta06} while the flux in the SW fluxes comes from the galaxy host. We also show the fits to an intrinsically blue QSO template, QSO1 from \citet{polletta06}, with a large A$_{V}$=$3-4$ based again on a Calzetti attenuation law. While this template fits worse than the QSO2, it is useful to illustrate the differences and it provides a way to estimate the bolometric luminosity of the QSO from the unobscured emission. We fit the QSO model in 3 steps. First, we do a coarse fit of the LW fluxes to the QSO2 template, then we fit all the photometry subtracting the best-fit fluxes from the QSO template with \texttt{Prospector} delayed-$\tau$ models, and lastly, we perform a simultaneous fit of the QSO template with galaxy SEDs drawn from the posterior of the \texttt{Prospector} fit. The results from this method are similar to those obtained with the modified version of FAST \citep{aird18} used in \citet{kocevski23}. The advantage of the \texttt{Prospector} fit is that it includes emission lines which can help shore up the limitations of the obscured QSO template that has a fixed set of emission lines. While this is not a fully self-consistent AGN method it helps to account for the contribution of emission lines to the photometry.

\begin{figure*}%[htp!]%[ht!]
\centering
\includegraphics[width=19cm,angle=0]{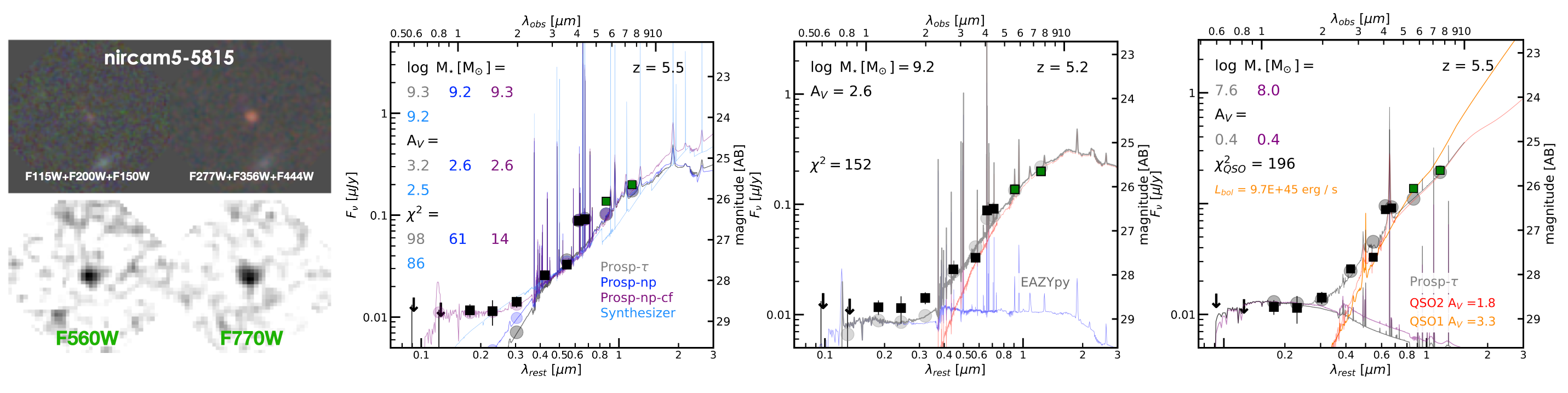}\\
\includegraphics[width=19cm,angle=0]{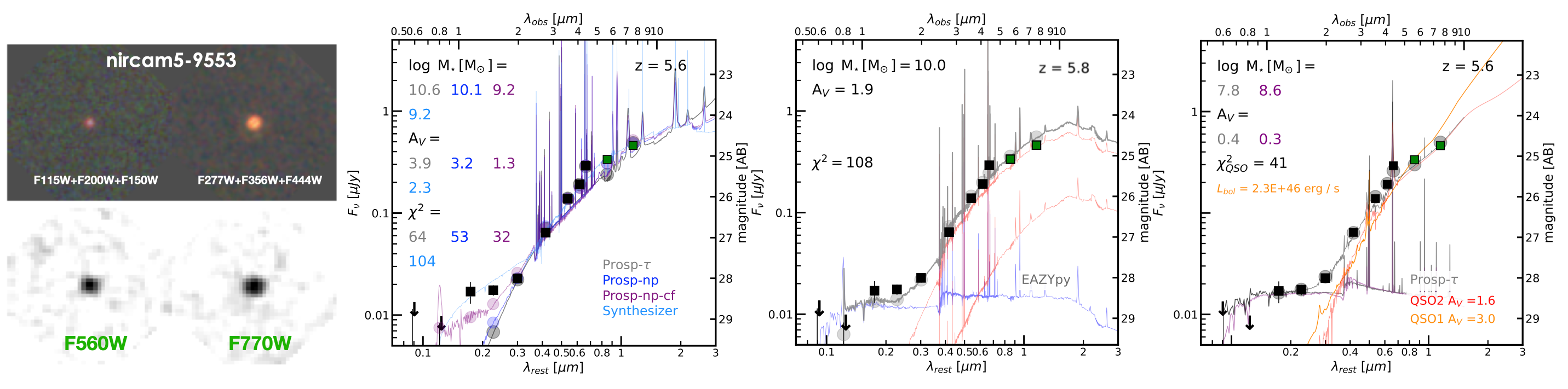}\\
\includegraphics[width=19cm,angle=0]{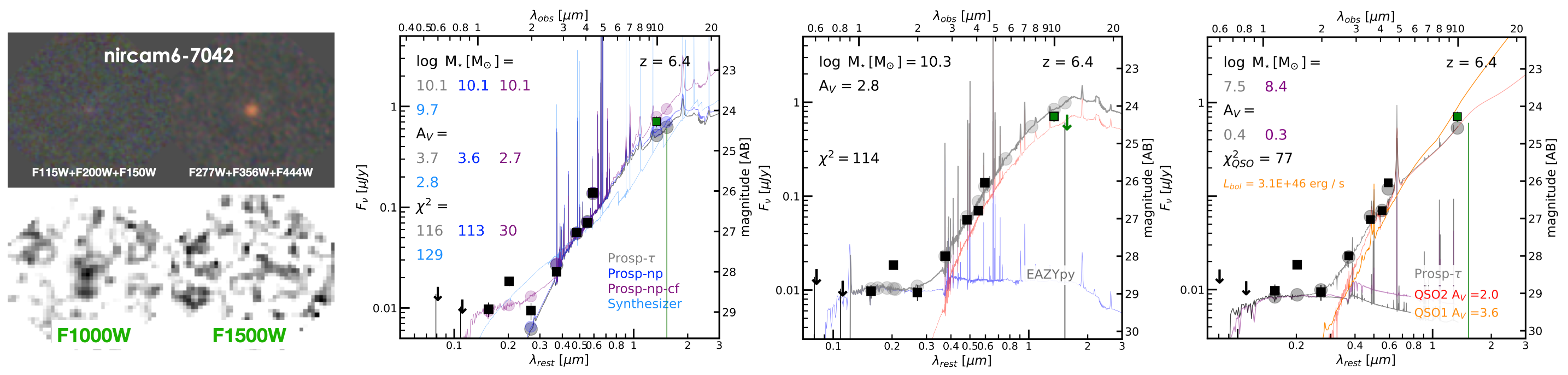}\\
\includegraphics[width=19cm,angle=0]{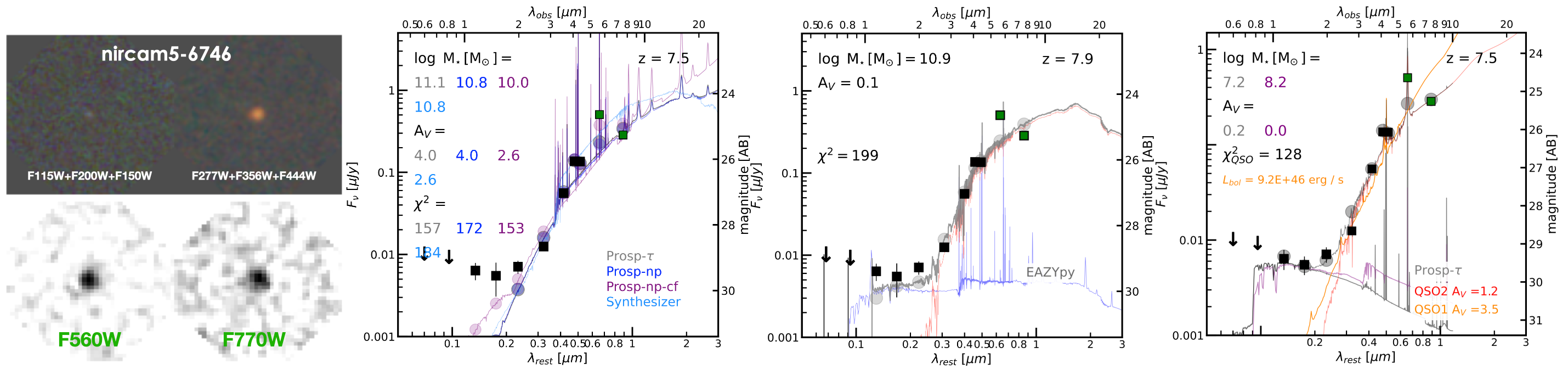}\\
\caption{Multi-band 2.5\arcsec$\times$2.5\arcsec\ cutouts of the MIRI detected EROs and best-fit SED models computed with \texttt{EAZYpy}, \texttt{Prospector}, \texttt{Synthesizer} and a hybrid of galaxy plus a red QSO2 template from \citet{polletta06}. The left panels illustrate that fits based on a single stellar population component provide a good fit to the overall LW NIRCam and MIRI photometry (black and green squares) but they systematically fail to reproduce the rest-frame UV probed by the SW NIRCam bands. The middle panels show that a composite model consisting of two (or more) stellar populations provides an excellent fit to all the bands by combining a red, massive, and dusty galaxy that fits the LW bands and a blue, low-mass galaxy that fits the SW bands but has little impact on the stellar mass. The right panels show that the hybrid of galaxy + QSO2 models provide an equally good (or better) fit to the SED than the other models. Here, a dust-obscured QSO dominates the LW photometry but does not contribute to the stellar mass of a blue unobscured host, and consequently leads to total stellar masses $\sim2$ orders of magnitude smaller than the other scenarios.}
\label{fig:seds2}
\end{figure*}

\begin{figure*}%[htp!]%[ht!]
\centering
\includegraphics[width=19cm,angle=0]{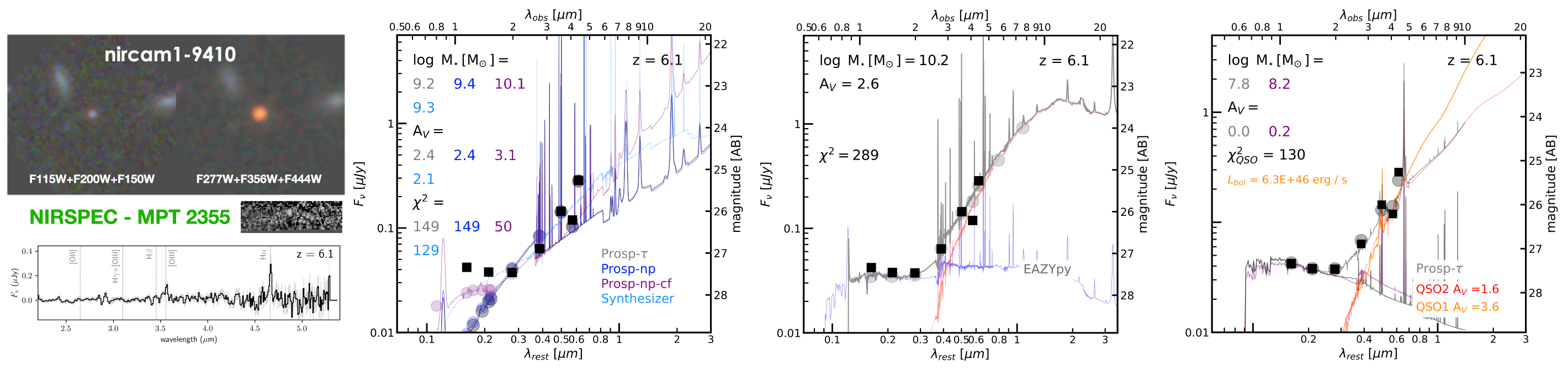}\\
\includegraphics[width=19cm,angle=0]{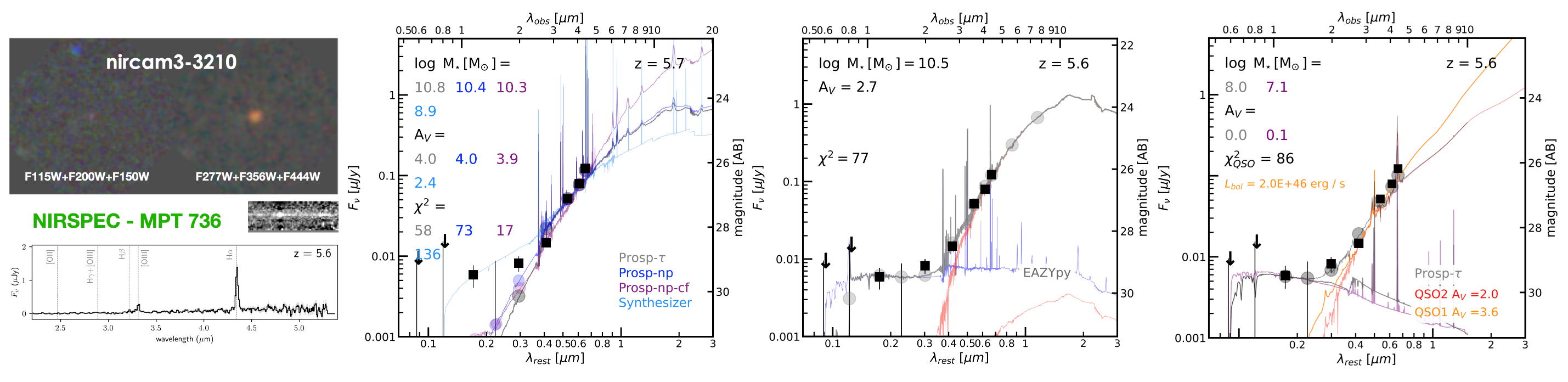}\\
\includegraphics[width=19cm,angle=0]{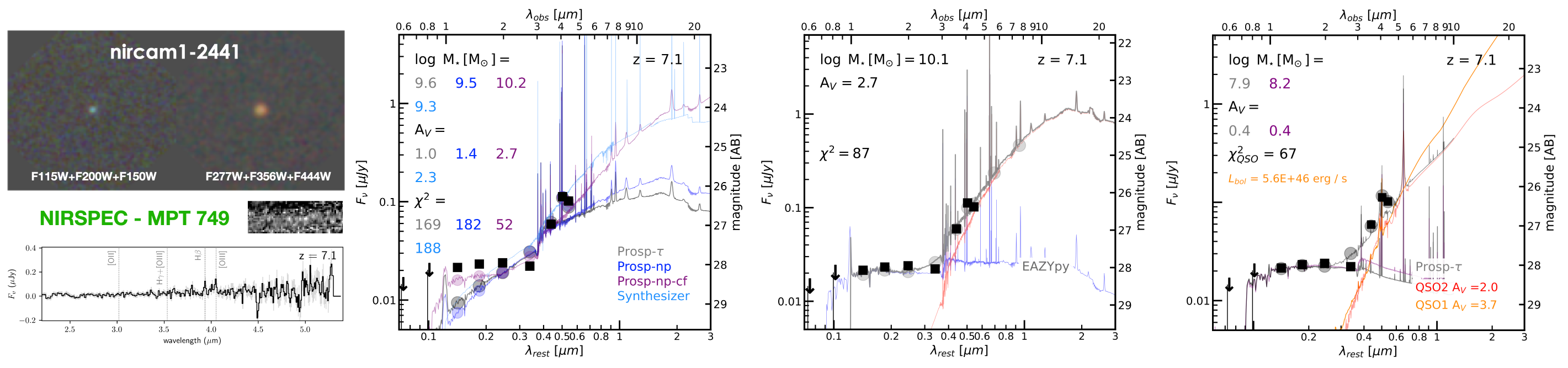}\\
\includegraphics[width=19cm,angle=0]{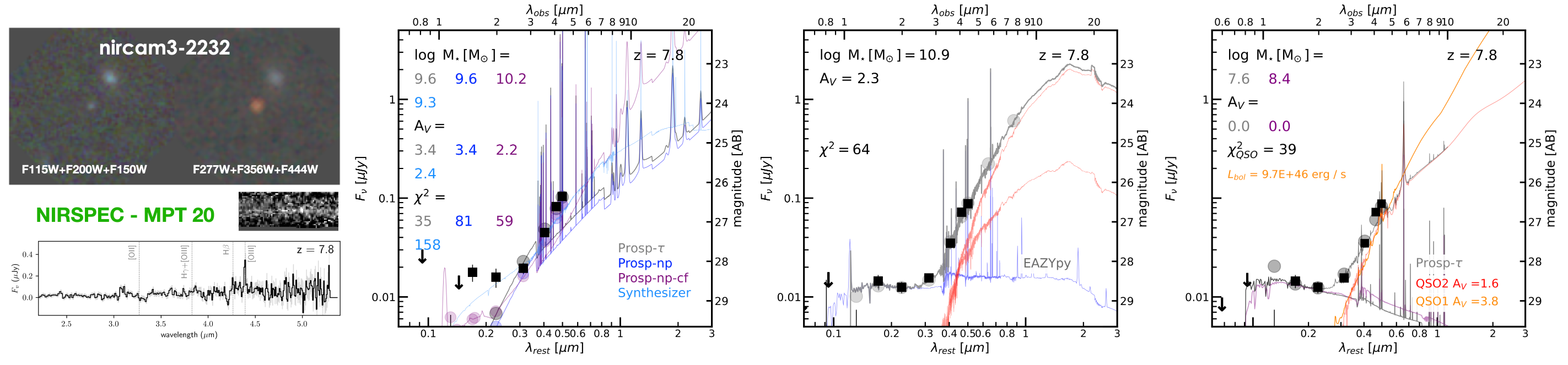}\\
\caption{Multi-band 2.5\arcsec$\times$2.5\arcsec\ cutouts and 2D/1D NIRSpec spectra of the NIRSpec detected EROs. The best-fit SED models computed with \texttt{EAZYpy}, \texttt{Prospector}, \texttt{Synthesizer} and a hybrid of galaxy plus the QSO2 template from \citet{polletta06} are the same as in Figure~\ref{fig:seds2} but fixed to the spectroscopic redshift.}
\label{fig:seds3}
\end{figure*}

\subsection{Photometric redshifts of the 4 EROs with MIRI detections}
\label{s:photozmiri}

The peculiar SEDs of the EROs and the high chances that some of the fluxes are at least partially boosted by emission lines make the photometric redshift estimates one of the key parameters and potentially one of the most problematic. For that reason, we run \texttt{EAZYpy} twice, first using the default modeling assumptions and a second time using the recently updated models which include a blue galaxy template with strong, high EW emission lines similar to those observed in recent NIRSpec spectra of $z>7$ galaxies. 
We also include in the analysis the redshift probability distributions (PDFz) the values computed in \citet{finkelstein23} using the original version of \texttt{EAZY} with an updated template set optimized for high redshift presented in \citet{Larson22}. The latter fits do not include the MIRI fluxes and thus allow us to gauge the impact of the additional photometry in the redshift likelihood. Lastly, we also include the PDFz estimate from the fiducial Prospector fit described in the previous section.

{\bf nircam5-5815:} The primary \texttt{EAZYpy} and \texttt{Prospector} solutions agree on a value of z$\sim5$ for which a strong \Ha\ emission would boost the fluxes in F410M and F444W. There is a secondary solution at z$\sim9$ for which the red F277W$-$F444W color is caused by a strong Balmer break. However, at that redshift, the galaxy should be an F150W drop-out, and the galaxy is clearly detected at $>5\sigma$. Therefore, we adopt the lower redshift solution as the primary. 

{\bf nircam5-9553:} The photometric redshift distributions from \texttt{EAZYpy} and prospector are quite consistent, peaking around $z\sim5.8$. At this redshift, the \OIII/\Hb\ and \Ha\ lines can contribute to the flux in F356W and F444W but not in F410M (or at least not significantly). There is a secondary peak at z$=8.7$ which also produces a relatively good fit. However, like in the previous galaxy that would require the F150W flux to be a drop-out, and the galaxy is faint but clearly detected in that band. Therefore we consider the low redshift solution as the primary.

{\bf nircam6-7042:} This is the only galaxy observed in the long wavelength MIRI bands. It has a faint but clear detection in F1000W but is not detected in F1500W. Similarly to the galaxies above, the PDFz exhibits a primary peak at $z=6.4$ and a secondary peak at $z\sim8.5$ which is closer to the value presented in \citet{labbe23}, $z=8.11$. The two different solutions try to fit an excess in F444W relative to F410M with a strong emission line, either \Ha\ or \OIII\ at low and high-z respectively. We notice however that the F277W flux for this source is above the relatively flat continuum delineated by the SW bands suggesting that it might be sampling the continuum redward of the 4000\AA\ break and therefore favoring the low-z solution. The F277W photometry in \citet{labbe23} appears to be fainter and closer to the bluer bands which might favor the high-z solution. At the redshift of the two possible solutions, the F1000W detection (and the upper limit in F1500W) still probes rest-frame wavelengths shorter than the 1.6~$\mu$m bump and thus cannot help discriminate between them.

{\bf nircam5-6746:} This galaxy presents a PDFz centered around $z=7-8$ with no secondary peaks at significantly different redshifts. The brighter MIRI flux in F560W relative to F770W also favors a redshift of $z=7.5$ suggesting that strong \Ha\ emission is boosting the flux in F560W and similarly \OIII\ in F410M and F444W. This galaxy is also discussed in \citet{akins23} with similar photometric redshift and consistent stellar population fits.

\begin{figure*}
\centering
\includegraphics[width=18.cm,angle=0]{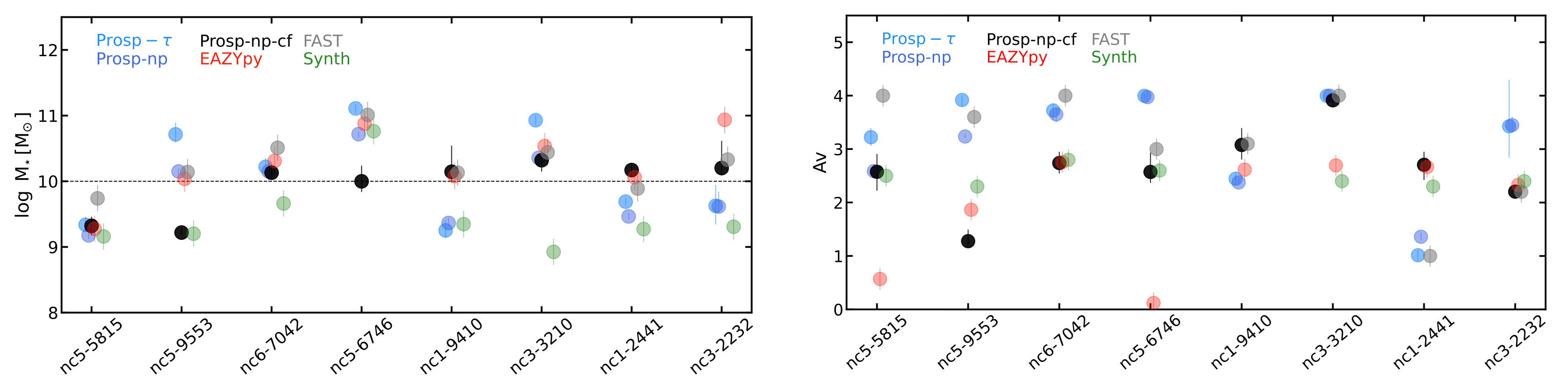}
\caption{Range in stellar masses and A$_{\rm V}$ obtained with different SED modeling assumptions for the 8 EROs with MIRI and NIRSpec observations. Overall, the values derived with the commonly used \texttt{EAZYpy} and \texttt{FAST} methods provide similar estimates as the fiducial 
\texttt{Prospector}-$\tau$ model, and they are typically the largest (red, grey, and light blue markers). The \texttt{Prospector}-np non-parametric model (dark blue) leads to smaller stellar masses by 0.4~dex, on average. A non-parametric model with a grey attenuation law, \texttt{Prospector}-np-cf (black), can lead to even smaller masses by 0.7~dex when MIRI fluxes are available, but it obtains similar values to the fiducial models where they are not. The stellar masses from \texttt{Synthesizer} (green) are the smallest by $\sim$1~dex relative to the fiducial values, but the accuracy of the fit is worse. The values obtained with the hybrid galaxy plus obscured QSO model (not shown) are much smaller, \lmass=$7-8$ because the QSO dominates the SED without contributing to the stellar mass of the blue, low-mass host.}
\label{fig:mass_range}
\end{figure*}

\subsection{Best fit properties and SEDs}
\label{s:bestfit}

Figure~\ref{fig:seds2} and \ref{fig:seds3} show the multi-band images, NIRSpec spectra, and SEDs for the 8 MIRI and NIRSpec detected galaxies jointly with the best-fit models obtained with the different codes outlined in the previous section. From left to right, the panels show the stellar population fits with \texttt{Prospector} ($\tau$-model and non-parametric) and \texttt{Synthesizer}, the composite stellar populations with \texttt{EAZYpy} (middle), and the hybrid galaxy+AGN models (right).

{\bf MIRI fluxes and the high EW emission line scenario:} The 4 galaxies with MIRI detections exhibit F560W and F777W fluxes that continue the red power-law trend outlined by the NIRCam LW bands. For 3 of them, the MIRI bands probe a spectral region redward of \Ha, which does not have any prominent emission lines. The exception is nc5-6746, at z$=7.5$, which seems to have an excess in F560W due to a strong \Ha\ line, but not in F777W, which also continues the same trend of increasingly larger fluxes as the other 3 galaxies. Therefore, the MIRI detections strongly suggest the presence of red continuum emission in these galaxies which disfavors the scenario where the red optical fluxes originate in a blue galaxy with very high EW emission lines masquerading as a red continuum. Nonetheless, we note that the best-fit SEDs for these EROs show strong emission lines and even emission line-driven excess in one or two of the LW NIRCam bands. However, these lines have relatively normal EWs for a massive star-forming galaxy ($\sim$100\AA) due to the presence of a red stellar continuum.

{\bf Prospector-$\tau$, -np and Synthesizer:} Overall, these models based on different SFHs but using the same \citet{calzetti} attenuation law provide a relatively good fit to the majority of the LW NIRCam bands and the MIRI fluxes. However, they all fail to reproduce the rest-frame UV fluxes probed by F115W, F150W, and, in some cases, F200W, regardless of the SFH. Both \texttt{Prospector} fits yield systematically lower fluxes in the rest-UV, while \texttt{Synthesizer} sometimes finds a trade-off between improving the fit to the UV bands at the expense of a worse fit to the optical bands. The reason behind this systematic issue for all the models is that the large dust attenuations required to reproduce the extremely red optical colors lead to even larger attenuations in the UV which completely suppress the predicted emission regardless of the stellar population parameters or SFHs; i.e., even non-parametric SFHs having substantial SFRs in the last 5 to 10~Myr still yield very red colors in the rest-frame UV. This problem is unavoidable for the typical attenuation laws such as Calzetti ($A_{2500}/A_{V}\sim2$), and it would be worse for steeper attenuation laws such as the SMC type  ($A_{2500}/A_{V}\sim2.6$) or a Milky-way type with a UV bump at 2175\AA. However, a shallower, greyer attenuation law, resulting perhaps from a more patchy distribution of the dust in the galaxy, could alleviate this problem.

{\bf Prospector-np-cf:} Indeed, the best-fit SED models derived with \texttt{Prospector} using non-parametric SFHs and a more complex, two-component dust attenuation model based on \citet{cf00} and \citet{kriek13} provide a better match to the UV fluxes with varying degrees of improvement. In this model, the diffuse attenuation is multiplied by a power law with index $n$ that increases/lowers the slope of the attenuation law relative to Calzetti (i.e., for $n=0$ it becomes Calzetti). The models that fit the UV fluxes best (e.g., nc5-5815, nc6-7042, or nc1-2441) all have similar attenuation laws which lean heavily towards the shallowest (greyest) possible attenuation law allowed by the priors ($n=0.4$, $A_{2500}/A_{\rm V}\sim1.4$); i.e, the posterior is not evenly sampled but rather skewed to the maximum value. The models without a significant improvement of the UV fit still return better $\chi^{2}$ than the Calzetti-based fits. For these galaxies differential attenuation between the stellar continuum and the emission lines introduced by the two-component \citet{cf00} prescription appears to allow stronger emission lines that improve the fit to the bands with emission line excesses. 

{\bf EAZYpy:} These models provide a good match to both the rest-UV and rest-optical SED. The difference with respect to the  \texttt{Prospector} and \texttt{Synthesizer} fits is that \texttt{EAZYpy} uses composite models that are linear combinations of templates with different ages, SFRs, and, crucially, dust attenuations. Consequently, the composite SED is not necessarily bounded by the same dust attenuation across the whole spectral range. The best-fit models for all the EROs are always a combination of at least two templates with very different properties: a young, blue galaxy with low dust attenuation that fits the relatively flat rest-frame UV emission, and an older galaxy, with large dust attenuation that fits the red optical emission.

{\bf Hybrid galaxy - red QSO:} The rightmost panels of Figure~\ref{fig:seds2} and \ref{fig:seds3} show the fits to the hybrid model of a blue galaxy and a dust-obscured QSO (QSO2 template in red). This model shows an excellent fit to the overall SED including the rest-UV and the MIRI fluxes. In this scenario, the continuum emission from the obscured QSO dominates the SED redward of F277W while the galaxy component dominates the rest-UV emission. Consequently, the best-fit galaxy model is a blue, low-extinction galaxy similar to the blue component in the \texttt{EAZYpy} composite. The grey and magenta lines in the fits illustrate the 1$\sigma$ range in the stellar masses which are, in all cases, very small \lmass$=$7-8. The main difference in the best-fit SEDs of QSO-dominated vs. galaxy-dominated scenarios is that in the latter, the stellar continuum typically exhibits a peak around $\sim1.6~\mu$m, whereas the QSO emission increases continuously toward the rest-frame mid-IR. Unfortunately, at $z>5$ the MIRI detections in F560W and F777W still probe rest-frame wavelengths shorter than 1.6~$\mu$m, and even for the one galaxy detected in F1000W, the rest-frame flux is still too close to 1.6~$\mu$m. Detections at longer wavelengths are clearly necessary to distinguish conclusively between a declining stellar continuum and rising QSO emission. The panels show also the fits using a blue QSO1 template with very large attenuations A$_{V}\gtrsim3$ (orange). These are generally a worse fit to the MIRI data because they have more steeply rising SEDs, but they help provide an order of magnitude estimate of the QSO bolometric luminosity. 

\begin{figure*}%[htp!]%[ht!]
\centering
\includegraphics[width=18cm,angle=0]{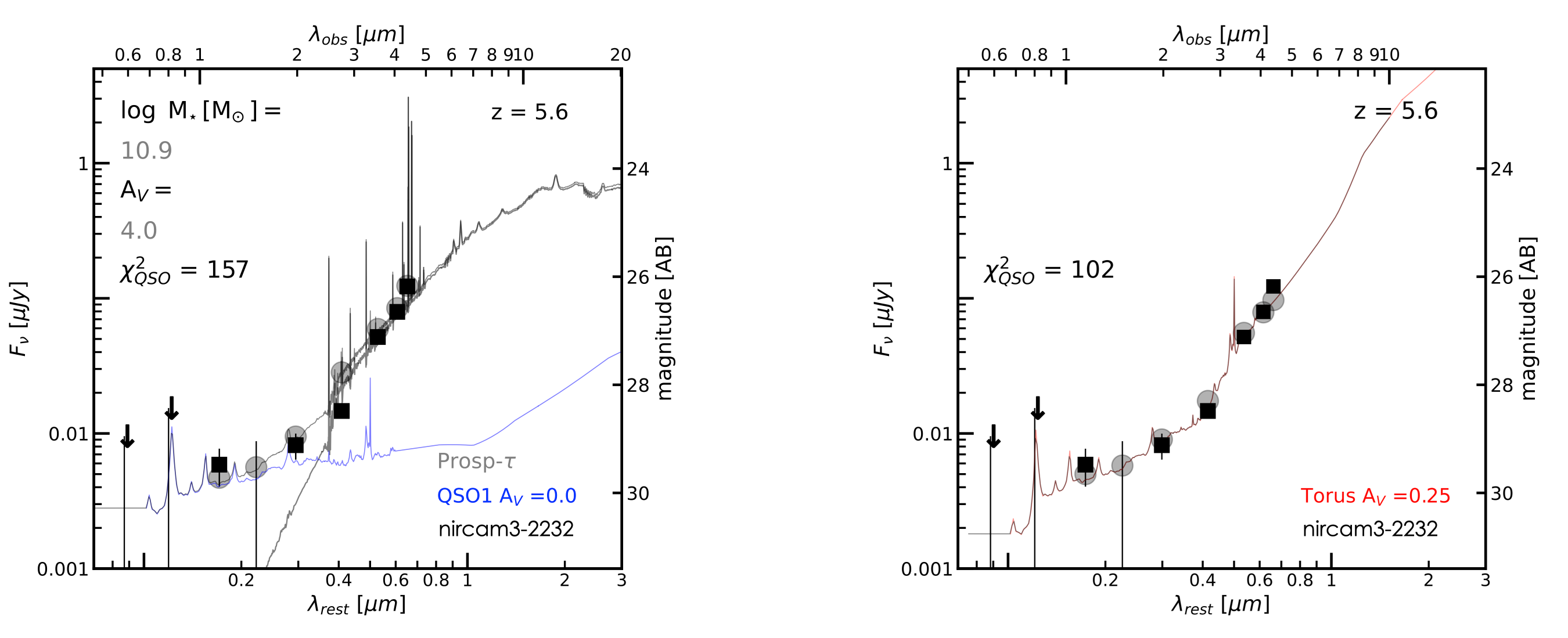}
\caption{Additional SED modeling scenarios involving a QSO. {\it Left:} a hybrid of dusty-galaxy dominated SED with a blue, low extinction QSO contributing only to the rest-UV emission. This scenario is similar to the \texttt{EAZYpy} fits replacing the blue galaxy with a blue QSO with a minimal impact on the stellar mass of the composite. {\it Right:} a pure QSO-dominated model based on the Torus template from \citet{polletta06} where the emission from the QSO outshines the galaxy host at all wavelengths. The intrinsic shape of the Torus SED is very similar to the bimodal SED of the EROs. However, we find that using a single template limits the flexibility of the fits and it leads to generally worse agreement ($\chi^{2}$) with the data.}
\label{fig:seds4}
\end{figure*}

{\bf Hybrid blue QSO - dusty galaxy, and pure QSO - Torus:} Figure~\ref{fig:seds4} shows the best-fit SED of nc3-3210 (the broad-line AGN with NIRSpec studied in \citealt{kocevski23}) to the other two possible scenarios involving a QSO: the QSO - torus emission, and the hybrid of a blue QSO and a red, dusty galaxy. In the torus model, the SED is completely QSO-dominated at all wavelengths (e.g., scattered UV light, attenuated optical emission, and mid-to-far IR re-emission by dust). Here we use the Torus template from \citet{polletta06} to fit the observed SED and we find that while the intrinsic shape of the Torus SED template is, to some extent, similar to the bimodal SED of the EROs, a single template is not flexible enough to obtain a better fit than any of the other scenarios discussed above. This is likely a limitation of our approach based on a single template, and it is possible that a more comprehensive AGN modeling code can fully reproduce the observed SED with higher accuracy. The scenario involving a galaxy plus a blue QSO is, to some extent, similar to the \texttt{EAZYpy} model. In both of them, the LW NIRCam bands are largely dominated by the emission of a red, dusty galaxy while the SW bands are dominated by a blue, low-extinction galaxy or QSO, respectively. Consequently, the inferred stellar masses and dust attenuations for the bulk of the galaxy are also very similar, since none of the blue components contribute significantly to the mass. 
These two scenarios are not discussed in detail for the other objects because, the blue QSO model leads to similar results for the stellar properties as the other galaxy-dominated scenarios, and the torus model does not provide constraints on the stellar mass of the host or the luminosity of the QSO.

\subsection{Stellar masses and attenuations}
\label{s:masses}

Figure~\ref{fig:mass_range} shows the range in stellar masses and dust attenuations for the 8 EROs obtained with the different SED modeling codes. We also include the stellar masses and attenuations computed with \texttt{FAST} and we use it as a benchmark model for the comparisons to study systematic effects. \texttt{FAST} has been widely tested in typical galaxies at low-to-mid redshift with accurate results, but it is critical to understand if there are potential issues modeling these high-z galaxies with peculiar SEDs.

The stellar masses computed with \texttt{FAST} and \texttt{EAZYpy} tend to be the largest, and they are very similar, with a median difference and scatter of $\Delta$\lmass (FAST-EAZYpy) $=$-0.01$\pm$0.3~dex. Although the SED fits with FAST do not reproduce the UV fluxes like the composite SEDs with EAZYpy, the effect on the stellar mass is very minor. This is because the red, dusty component in the EAZYpy fit, which is similar to the overall FAST fit, dominates the stellar mass over the young, blue component which has a much lower mass-to-light ratio.

Interestingly, the median difference with respect to the stellar masses computed with the fiducial \texttt{Prospector} fits ($\tau$-model with a Calzetti attenuation) is relatively small, with a larger scatter $\Delta$ \lmass(FAST-Prospector-$\tau$)$=$-0.16$\pm$0.49~dex.  This means that despite the more flexible modeling of key parameters like emission line strength or metallicity, the stellar mass is mostly driven by the need to fit the red optical slope with a high dust attenuation.  In fact, the cases where the \texttt{Prospector} fits obtain the largest stellar masses are typically those where the extinction is maximum Av$\sim$4. Note also that the extinction values from \texttt{FAST} and \texttt{Prospector} are typically the largest, ranging between Av$=3-4$. The \texttt{EAZYpy} fits have lower extinctions Av$=2-3$ in part because of the combination with a blue template (A$_{\rm V}=0$), but sometimes because it includes a red, quiescent template which also has a low attenuation but a large mass-to-light ratio, which, in turn, leads to larger stellar masses (e.g., as in nc3-2232).

The \texttt{Prospector} fits with non-parametric SFHs capped at a maximum formation age of 100~Myr and a Calzetti attenuation law leads to systematically lower stellar masses than the fiducial \texttt{Prospector}-$\tau$, $\Delta$\lmass($\tau$-np)=-0.23$\pm$0.21~dex. This is because the fiducial model has a maximally old start of the SFH at 90\% of the age of the Universe at the redshift of the galaxy and, consequently, tends to form more stars over a longer period of time. Consequently, the masses are even smaller relative to \texttt{FAST} $\Delta$\lmass(FAST-Prospector-np)$=$-0.39$\pm$0.26~dex.

The \texttt{Prospector} fits with non-parametric SFH and a more flexible attenuation law based on \citet{cf00}, which provides the best SED fits, exhibit an interesting behavior. For the 4 galaxies with MIRI detections, the stellar masses are significantly lower $\Delta$\lmass (FAST-Prospector-np-cf)$=$-0.68$\pm$0.28~dex, but for the 4 galaxies with NIRSpec the difference is nearly zero, $\Delta$\lmass (FAST-Prospector-np)$=$-0.01$\pm$0.16~dex. The reason for this difference is clearly visible in the SED fits of Figure~\ref{fig:seds2} and \ref{fig:seds3}. Without MIRI data to constrain the continuum emission beyond F444W, \texttt{Prospector} favors solutions with a stronger continuum (i.e., more massive) and lower EWs for the lines. For example, in nc3-3210 or nc1-9410, the best-fit models with \texttt{Prospector}-$\tau$ vs. \texttt{Prospector}-np-cf would exhibit differences in the predicted F560W and F770W fluxes of the order of 1 to 1.5~mag.

The fits with \texttt{Synthesizer} provide the smallest stellar mass estimates, nearly 1~dex smaller than FAST, $\Delta$\lmass (FAST-Synthesizer)$=$-0.98$\pm$0.33~dex. As discussed in the previous section, these SED fits are, overall, less accurate than the other codes, but tend to fit the UV region a bit better at the expense of a worse fit to the optical. As a result, they have lower attenuations A$_{\rm V}\sim$2~mag and, consequently, lower stellar masses.

Lastly, in the hybrid galaxy plus obscured QSO scenario the latter completely dominates the bulk of the emission in the LW bands. However, it does not contribute to the stellar mass, which depends exclusively on the faint blue galaxy host. Consequently, inferred stellar masses are $\sim2$ orders of magnitude, \lmass=$7-8$, smaller than in any of the scenarios where the bright LW continuum originates in a dusty massive galaxy.

In summary, the commonly used methods based on $\tau$-models and a Calzetti attenuation, or variants of \texttt{EAZY} with the default templates (including the reddest dusty/old templates) are likely to obtain the largest stellar masses. Non-parametric or similar SFHs that limit the age of the galaxy to relatively young values ($\lesssim$100~Myr) lead to -0.4~dex smaller stellar masses. The addition of a more flexible dust modeling to allow greyer attenuation curves can lead to stellar masses up to 0.7~dex smaller. However,  without MIRI data, the stellar masses can also be as high as for the fiducial $\tau$-models.

%include{table1}
\begin{table*}
\scriptsize
%\small
\begin{center}
\caption{MIRI and NIRSpec EROs at $5<z<9$}
\label{tab:objects}
\begin{tabular}{p{1.65cm}lcccp{1.1cm}cp{1.1cm}cp{1.1cm}cp{1.1cm}cp{1.1cm}cp{1.1cm}cp{1.1cm}cp{1.1cm}c}
%{lcccccccccc}
\hline\hline
 ID  &  R.A.  &  Dec.  &  $z_{\rm phot}$  &  $z_{\rm spec}$  
 & $\log$M$_{\star}$  &  $\log$M$_{\star}$ & $\log$M$_{\star}$  &  $\log$M$_{\star}$ &  $\log$M$_{\star}$ &  $\log$M$_{\star}$ &  $\log$M$_{\star}$ \\
   &   (deg)  &  (deg)  &    &    &  [M$_{\odot}$]  &  [M$_{\odot}$]  &  [M$_{\odot}$] & [M$_{\odot}$]  &[M$_{\odot}$]  & [M$_{\odot}$]  & [M$_{\odot}$] \\
(1)  &  (2)  &  (3)  &  (4)  &  (5)  &  (6)  &  (7)  &  (8)  &  (9) & (10) & (11) & (12) \\
 \hline
nc5-5815                            & 214.975531  &  52.925267  &  $5.5^{+0.9}_{-0.1}$  &  ---        &  9.34$^{+0.12}_{-0.10}$  & 9.18$^{+0.07}_{-0.06}$  & 9.32$^{+0.14}_{-0.14}$         & 9.28$\pm0.20$ &  9.74$\pm0.20$  & 9.16$^{+0.03}_{-0.03}$   & 7.80$^{+0.20}_{-0.20}$   \\
nc5-9553                            & 214.990983  &  52.916521  &  $5.6^{+0.3}_{-0.3}$  &  ---        & 10.71$^{+0.18}_{-0.12}$  &10.15$^{+0.04}_{-0.05}$  & 9.22$^{+0.08}_{-0.04}$         &10.04$\pm0.20$ & 10.14$\pm0.20$  & 9.20$^{+0.04}_{-0.04}$   & 8.20$^{+0.40}_{-0.40}$   \\
nc6-7042                            & 214.840543  &  52.817942  &  $6.4^{+0.7}_{-0.2}$  &  ---        & 10.22$^{+0.13}_{-0.14}$  &10.15$^{+0.13}_{-0.15}$  &10.13$^{+0.11}_{-0.10}$         &10.31$\pm0.20$ & 10.51$\pm0.20$  & 9.66$^{+0.06}_{-0.06}$   & 7.95$^{+0.45}_{-0.45}$   \\
nc5-6746$^{\dagger}$                & 214.990983  &  52.916521  &  $7.5^{+0.4}_{-0.2}$  &  ---        & 11.11$^{+0.06}_{-0.07}$  &10.72$^{+0.04}_{-0.06}$  &10.00$^{+0.24}_{-0.16}$         &10.88$\pm0.20$ & 11.01$\pm0.20$  &10.76$^{+0.12}_{-0.12}$   & 7.70$^{+0.50}_{-0.50}$   \\
nc1-9410                            &  215.008490  &  52.977971  & $6.3^{+0.1}_{-0.2}$  &  $6.132$    &  9.25$^{+0.16}_{-0.11}$  & 9.37$^{+0.08}_{-0.05}$  &10.14$^{+0.40}_{-0.16}$         &10.08$\pm0.20$ & 10.13$\pm0.20$  & 9.35$^{+0.05}_{-0.05}$   & 8.30$^{+0.10}_{-0.10}$   \\
nc3-3210$^{\dagger\dagger}$         &  214.809145  &  52.868482  & $5.7^{+1.5}_{-0.1}$  &  $5.614$    & 10.93$^{+0.08}_{-0.08}$  &10.36$^{+0.04}_{-0.06}$  &10.32$^{+0.12}_{-0.17}$         &10.54$\pm0.20$ & 10.44$\pm0.20$  & 8.93$^{+0.06}_{-0.06}$   & 7.55$^{+0.45}_{-0.45}$   \\
nc1-2441                            &  215.002842  &  53.007588  & $7.6^{+0.5}_{-0.2}$  &  $7.092$    &  9.69$^{+0.06}_{-0.06}$  & 9.47$^{+0.05}_{-0.05}$  &10.17$^{+0.11}_{-0.11}$         &10.06$\pm0.20$ &  9.89$\pm0.20$  & 9.27$^{+0.10}_{-0.10}$   & 8.35$^{+0.15}_{-0.15}$   \\
nc3-2232$^{\dagger\dagger\dagger}$  &  214.830687  &  52.887769  & $8.6^{+0.5}_{-0.7}$  &  $7.769$    &  9.63$^{+0.32}_{-0.29}$  & 9.62$^{+0.23}_{-0.10}$  &10.20$^{+0.41}_{-0.11}$         &10.94$\pm0.20$ & 10.33$\pm0.20$  & 9.31$^{+0.08}_{-0.08}$   & 7.90$^{+0.10}_{-0.10}$   \\
\hline \hline
\end{tabular}
\end{center}
\tablecomments{
(1) Source ID in the CEERS catalog.
(2) Right ascension (J2000).
(3) Declination (J2000).
(4) Photometric redshift in \S~\ref{s:photoz}
(5) Spectroscopic redshift from NIRSpec. See decription in  \S~\ref{s:stellarpop} for the following:
(6) Stellar masses derived using \texttt{Prospector}-$\tau$
(7) Stellar masses derived using \texttt{Prospector}-np
(8) Stellar masses derived using \texttt{Prospector}-np-cf
(9) Stellar masses derived using \texttt{EAZYpy}
(10) Stellar masses derived using \texttt{FAST}
(11) Stellar masses derived using \texttt{Synthesizer}
(12) Stellar masses derived using the hybrid galaxy - obscured QSO model.
$\dagger$ : This object is also studied in \citet{akins23}.
$\dagger\dagger$ : This object is also studied in \citet{kocevski23} as MPTID-746
$\dagger\dagger\dagger$ : This object is also studied in \citet{fujimoto23} as MPTID-20
}
\end{table*}

\section{Discussion}
\label{s:discussion}
\subsection{Likelihood of the dusty galaxy scenario}

In the previous sections, we discussed 3 possible scenarios where a dusty, star-forming galaxy can fit the overall SED of the EROs dominating the emission in the rest-frame optical: with a flat, grey attenuation law or with a secondary component which is either a blue, low-extinction galaxy or a blue QSO, that fits rest-frame UV.

Looking at these possibilities in the light of the point-like, unresolved sizes for all these galaxies the scenario with two distinct stellar components seems quite unlikely. Such a model would make more sense on an extended galaxy with clearly differentiated regions (e.g., clumps, or a bulge).  On the other hand, a compact size might help explain the very grey attenuation law in terms of the geometry and distribution of dust in a high-density environment. For example, rather than a dust shell scenario we might have a mixed star-dust distribution (probably clumpy) which produces gray attenuation laws including huge extinctions ($A_{V}\gtrsim$20 mag or more), but also significant scattering resulting in much lower and gray total attenuations and, consequently, bluer UV colors \citep{wg00}.

The scenario involving a blue, low-luminosity QSO is also plausible and it can help explain why the colors of these EROs are very different from those of F150W EROs and other dusty galaxies at higher redshift recently identified with JWST (e.g., \citealt{zavala23}, \citealt{pgp23a}), which are red in all the NIRCam bands.  As discussed in \citet{kocevski23} (also \citealt{fujimoto22}) this scenario could be a transitional phase, in the evolution of a dust-obscured starburst that is clearing up the dust and leading the way to an unobscured QSO. Note that while bluer UV colors have been reported in dusty star-forming galaxies at $z\gtrsim3$ with large IR-luminosities (e.g., \citealt{casey14}), these EROs are very blue, with a relatively flat UV continuum in f$_{\nu}$ that implies very steep UV-slopes, $\beta\lesssim-2$ for the high attenuations implied by the SED modeling, A$_{V}>3$~mag.

Taken together, the different colors and morphologies of these EROs relative to the other massive dusty galaxies might be an indication that they are a distinct population perhaps undergoing a strong nuclear starburst phase as seen for example in some of the radio/sub-mm detected galaxies at $z>3$ (\citealt{tadaki17a}; \citealt{barro17}). To some degree, this scenario might be similar to that of the compact star-forming galaxies at $z\gtrsim2-3$ which are also small (but resolved, r$_{\rm e}\sim$1~kpc), massive, and dusty (e.g., \citealt{barro13}; \citealt{williams14}; \citealt{nelson14}; \citealt{dokkum15}), and exhibit a large fraction of X-ray AGN detections (\citealt{kocevski17}). Indeed, galaxy formation models suggest that the progenitors of those compact SFGs could be even smaller at higher redshift due to the larger gas reservoirs leading to wet-compaction events that result in the formation of a very dense core (e.g., \citealt{zolotov15}; \citealt{wellons15}; \citealt{tacchella16}). Nevertheless, it seems odd that all these EROs at $z=5-9$ are unresolved. If they were to evolve into compact SFGs at $z\lesssim3$ we would expect some of them to be transitioning from purely unresolved to the characteristic mass-size relation that compact SFGs follow at z$\gtrsim2-3$ \citep{barro17}.

\subsection{Likelihood of the obscured AGN scenarios}

%AGN
An alternative scenario to the dusty star-forming galaxies where we expect unresolved, point-like sources and peculiar, non-stellar SEDs is in AGNs where a bright QSO can outshine the emission of its host in different spectral ranges from the UV to the mid-IR. For example, hybrid galaxy + AGN SEDs where the latter dominates the near-to-mid IR emission are a relatively common occurrence in galaxy surveys at mid-to-high redshifts (\citealt{stern05}; \citealt{lacy07}; \citealt{donley12, donley18}). In the previous sections, we discussed 2 possible scenarios where an obscured AGN can fit the overall SED of the EROs dominating the red optical emission: 1) combined with a blue, low-mass galaxy host or in a pure AGN model where the emission from the QSO dominates at all wavelengths.

The first scenario would imply that all these EROs are low-mass galaxies whose optical to IR fluxes are completely outshined by the emission of an obscured QSO. The limiting factor in this scenario is the bolometric luminosity and implied black-hole mass of the QSOs which should be at least one or two orders of magnitude lower than the stellar masses of the hosts (e.g., \citealt{kormendy13}). The stellar masses of the blue, low-extinction hosts inferred in the previous section range between \lmass$=7-8.5$. Therefore, we would expect black hole masses of the order of \lmass$=6-7$ and, based on the typical luminosity-black hole relation \citep{greene07}, QSO bolometric luminosities of $L_{\rm bol}\sim10^{44-45}$erg~s$^{-1}$ or smaller, since this is the value at the higher end of the accretion rate, L$_{\rm bol}$/L$_{\rm Edd}$$=$1.

Unfortunately, the estimate of the bolometric luminosity of an obscured QSO requires X-ray, UV, or bolometric luminosities, none of which can be easily computed for these galaxies. For intrinsically blue QSOs the total luminosities can be estimated from monochromatic luminosities using bolometric corrections (e.g., \citealt{richards06}). However, for obscured AGNs, the total luminosities are usually inferred from rest-frame IR luminosities of the total IR luminosity (e.g., \citealt{donley12}; \citealt{runnoe12}), which for these galaxies would require MIRI fluxes at the longest wavelengths. Thus, the only alternative to estimate a luminosity is to fit the SED with a blue QSO template heavily obscured with a Calzetti attenuation law and then transform the dust-corrected UV luminosity into L$_{\rm bol}$ (e.g., L$_{\rm bol}$=5.15L$_{3000}$, \citealt{richards06}). The values obtained for the EROs with this method range between $L_{\rm bol}\sim10^{46-47}$erg~s$^{-1}$ which are 1 to 2 dex larger than the expectation from typical low redshift BH mass to stellar mass ratios (i.e., they would be very luminous QSO). We caution however, that this estimate is a large oversimplification since, as shown in \S~\ref{s:stellarpop}, the red QSO SED (QSO2) differs from the attenuated blue QSO SED (QSO1+Calzetti). The true attenuation law of an obscured QSO depends on multiple factors such as the geometry and distribution of dust in the torus or the line-of-sight inclination. In compact galaxies at high-z, it might even depend on the galaxy-wide conditions (gas/dust fractions; e.g., \citealt{gilli14}). Consequently, the bolometric luminosities of obscured QSOs are probably lower than the values estimated with a blue QSO template, which should be considered upper limits.

In the second scenario, the obscured QSO completely outshines the galaxy host emission across the whole spectral range; i.e., both the SW and LW NIRCam fluxes arise from the QSO. This scenario would be the most plausible based on the unresolved sizes of these galaxies in all the NIRCam bands. Unfortunately, a more detailed characterization of the bolometric luminosity in this type of scenario requires more complex modeling of the extinction and scattering of the QSO emission that is beyond the scope of this paper. Interestingly, in this scenario, the constraints on the bolometric luminosity of the QSO might be less strict since the galaxy host does not have to be detected in the UV. Therefore, a slightly more massive and dusty galaxy, \lmass$=8-9$, can perhaps hide under the bright red continuum of the QSO without having a significant impact on the observed SED.

\begin{figure}%[htp!]%[ht!]
\includegraphics[width=8.7cm,angle=0]{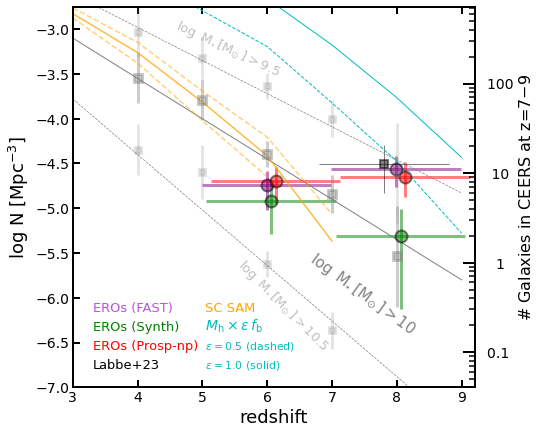}
\caption{Galaxy number densities with stellar masses above \lmass$>10$  (dark grey, \lmass$>$9.5 and 10.5 in shaded grey) as a function of redshift computed from SMFs in the literature (see text). The purple, green, and red markers indicate the number density of EROs with \lmass$>10$ based on three different modeling scenarios which typically encompass the minimum/maximum stellar mass estimates (see \S\ref{s:masses}). The orange lines show a similar prediction derived from median and percentiles of 100 CEERS-sized draws of a 2-deg$^2$ lightcone based on the Santa Cruz SAMs. Similarly, the cyan lines show the predictions from mock lightcones with larger baryon conversion efficiencies ($\epsilon=0.5$ and 1). At $z=5-7$ the density of EROs is lower than the values from the literature, however, at $z>7$ the density can be up to a factor of $\sim$10 larger for some of the estimates with the largest stellar masses. While this difference can still be reconciled with the large uncertainties from the SMFs and the SED modeling variations, the discrepancy at larger masses, \lmass$>$10.5, is much larger. We expect 1 galaxy in an area 10 to 20 times larger than CEERS and we find 3.}
\label{fig:density}
\end{figure}

\subsection{Implications for number densities, mass/luminosity functions}

As discussed in \citet{labbe23}, if all of these EROs are dusty galaxies with relatively large masses \lmass$\sim 9-10$, or up to \lmass$\sim$11 for some of the most extreme objects, their number densities can lead to some tension with the observed stellar mass functions and would imply higher than expected star formation efficiencies \citep{boylan23}. We review the number density estimates using the full sample of 37 EROs selected over the larger area of the full CEERS survey and spanning a broader redshift range from $z=5-9$. Figure~\ref{fig:density} shows the redshift evolution in the number density of galaxies with stellar masses larger than \lmass$=10$ (9.5 and 10.5 in dashed lines) derived from pre-JWST stellar mass functions (SMFs) in the literature (\citealt{muzzin13smf}; \citealt{grazian15}; \citealt{stefanon15}; \citealt{song16}; \citealt{stefanon21}). The orange lines show a similar prediction from mock catalogs based on the Santa Cruz semi-analytic models \citep{somerville2015,Somerville2021,Yung2019b,Yung2022}, which shows the median and 84th and 16th percentiles from 100 CEERS-sized fields subsampled from a 2-deg$^2$ lightcone \citep{Yung2023} to illustrate the effect of cosmic variance. These results have been shown to agree well with observed luminosity functions and other observations in this redshift range \citep{Yung2019a, Yung2019b}.  The purple, red, and green markers show the number densities of EROs with \lmass$>10$ at redshifts z$=5-7$ and z$=7-9$, estimated with \texttt{FAST}, \texttt{Prospector-np} and \texttt{Synthesizer}. The error bars indicate Poissonian errors. As discussed in the previous sections, these values generally bracket the largest to smallest stellar mass estimates and therefore provide a way to estimate the impact of the SED modeling choices on the number densities.  

The densities of EROs with \lmass$>10$ at z$=5-7$ are all slightly under the values from the literature which still allows additional, non-ERO, massive galaxies to exist at this redshift (e.g., \citealt{zavala23}) without tension with the literature.  At $z\sim7$, the expected number of galaxies with \lmass$>10$ in the area of CEERS is roughly one (with large errors) while the densities of EROs, inferred from the different stellar mass estimates, range between 2 and 10. Nonetheless, these differences are still within the range of the uncertainties in the SED modeling and cosmic variance. Furthermore, it is possible that pre-JWST stellar mass functions missed some of the most massive galaxies. At the largest masses \lmass$>10.5$, however, the difference increases up to a factor of $\sim60$. We should not detect any such galaxies in the area of CEERS (or even an area 10 times larger). \citet{labbe23} reported one of those galaxies in their sample. Here we identify 5 galaxies (including Labb\'{e}'s) with masses above \lmass$>10.5$ by at least 2 out of the 3 estimates, 2 of them at $z=5-7$, and another 3 at $z=7-9$ (nc1-10084, nc5-3637 and nc8-13596, see Table~\ref{tab:otherobjects}). These galaxies are also among the brightest in F444W$\sim$23 ~mag (up to 2 mag brighter than the median of all EROs) which indicates that they are different in some way and perhaps they are the ones that are AGNs. Nonetheless, an individual analysis of these sources carefully characterizing their photo-z and masses is required to clarify the strong discrepancy with respect to the expected densities. In summary, while the number densities of EROs exhibit some tension with the predictions of pre-JWST stellar mass functions if they are all dusty galaxies, the numbers match relatively well if all the masses are closer to \lmass$=10$, as predicted by some of the SED modeling scenarios. Nonetheless, even if just a few of them are confirmed to be very massive, \lmass$>10.5$, the discrepancy with the SMFs would be very large.

Comparing to simulations, the cyan lines in Figure~\ref{fig:density} show the predictions based on mock lightcones presented in \citet{Yung2023}, for which dark matter halos have been extracted from N-body simulations in a standard $\Lambda$CDM cosmology. Each line indicates the density of objects that would result if each halo is able to convert a different fraction of its baryon content into stars (i.e. $m_* = \epsilon f_b M_{\rm halo}$), $\epsilon=0.5$ and 1, respectively. A value of $\epsilon=1$ is expected to yield an extreme upper limit since the fractions in the local Universe are typically less than $\epsilon \sim 0.2$. Models based on similarly low efficiencies, such as the Santa Cruz SAM (orange lines) yield a good agreement with the density of EROs at $z=5-7$. However, the implied masses of the EROs in the $z=7-9$ bin, if they are primarily powered by stars, would imply significantly higher than expected values of this baryon conversion efficiency $\epsilon$, although not in fundamental tension with $\Lambda$CDM.

The implications for the number densities and luminosity functions in the obscured AGN scenarios are more complicated since these estimates depend on the bolometric luminosities which, as mentioned above, are particularly complicated for obscured AGNs. Nevertheless, if we again use the upper limits estimated with the blue QSO template and we assume that all EROs are obscured AGNs, we infer a large number density of luminous (but obscured) QSO, 4.2$\times$10$^{-5}$ and 3.9$\times$10$^{-5}$ Mpc$^{-3}$, at $z=5-7$ and $z=7-9$, respectively. These values are more than two orders of magnitude larger than the expected number of bright, intrinsically blue QSO at z$\sim6$ ($\sim$10$^{-8}$~Mpc$^{-3}$; \citealt{matsuoka18}). However, if the luminosities were a factor of 5$-$10 lower, the density would be closer to that of the less luminous, X-ray selected AGNs ($\sim$10$^{-6}$~Mpc$^{-3}$; \citealt{giallongo19}), or even to the rapidly increasing number of new AGNs identified with JWST/NIRSpec based on their broad emission lines (\citealt{kocevski23}; \citealt{larson23}; \citealt{harikane23a}), including one of the EROs discussed in this paper. A significant increase in the number of AGNs at $z>7$ relative to lower redshifts could indicate that we are probing a key era of very quick supermassive black hole growth and short duty cycles occurring in the first 1~Gyr of the Universe.

\subsection{Prospects for revealing the nature of these EROs}

These sources are complex to interpret, even though the short wavelength photometry with MIRI and the spectroscopy with NIRSpec help place better constraints on the presence of a red continuum or the redshift of these sources, it is not enough to break the degeneracies in the possible modeling scenarios. Additional MIRI photometry at longer wavelengths can distinguish between the rising continuum of an obscured QSO and the decline in the stellar SED past 1.6~$\mu$m. Deeper NIRSpec spectroscopy of their rest-frame UV or rest-frame optical can reveal high excitation emission lines (e.g., C~II and Mg~II, or He~II and [Ne~V]), indicative of AGNs or reach the stellar continuum showing absorption lines that would confirm the presence of an underlying stellar population.

\section{Summary}
\label{s:summary}

We identify 37 extremely red objects (EROs) in the CEERS field with NIRCam colors F277W$-$F444W$>1.5$ mag, down to a limiting magnitude of F444W$<28$ mag. These are candidate massive dusty galaxies at z$>$5.

\begin{itemize}
\item A key defining feature of these EROs is that {\it all} of them have blue colors in the SW NIRCam bands (F150W$-$F277W$\sim$0). The color difference in the SW and LW bands indicates that these  galaxies have bimodal SEDs consisting of a red, power-law slope  ($\alpha_{\nu}>$3) in the rest-frame optical, and a blue, flat slope in the rest-frame UV. These colors and SEDs are very different from those of other EROs or massive dusty galaxies at lower or similar redshifts.

\item Another key feature is that {\it all} of them are remarkably compact and featureless. The light profile fits with \texttt{GALFIT} indicate that they are unresolved, point-like sources in all the NIRCam bands. This differs again from the typical spread in stellar mass-size of other EROs or massive galaxies at similar redshifts.
  
\item Their photometric redshifts, stellar masses, and dust extinctions derived with standard SED fitting codes \texttt{EAZYpy} and \texttt{FAST} range from $5<z<9$ with median values $<z>=6.9^{+1.0}_{-1.6}$, \lmass$=10.2^{+0.5}_{-0.4}$ and A$_{V}=3.0^{+1.3}_{-1.2}$. However, if the red colors are not due to stellar continuum emission in a dusty galaxy, these values might be overestimated. Alternative scenarios include: emission lines with extreme EWs$>1000$\AA\ from a galaxy or AGN boosting the LW fluxes, a hybrid of galaxy and a dusty QSO with the latter dominating the LW continuum, or an AGN dominating the whole SED.
    
\item Four of these EROs within the limited MIRI overlap with the CEERS/NIRCam mosaic are clearly detected showing that the extremely red colors extend to longer wavelengths. Another 4 EROs were observed with NIRSpec and they exhibit \OIII\ and \Ha\ emission lines which confirm spectroscopic redshifts in the $z=5-9$ range. The MIRI detections at rest wavelengths redward of the most prominent emission lines, indicate the presence of a continuum and disfavors a scenario where these EROs are intrinsically blue galaxies with high EW emission lines masquerading as a red continuum.
        
\item We investigate the likelihood and implications of the different modeling scenarios using the 8 MIRI and NIRSpec-detected EROs to test a variety of codes with flexible options to characterize the stellar continuum, emission lines, dust attenuation, SFHs, etc. For scenarios where the LW bands are dominated by a dusty galaxy, we find: 1) SED models based on either parametric or non-parametric SFHs and a Calzetti attenuation law fail to reproduce the blue, rest-frame UV emission regardless of the modeling assumptions (age, metallicity, etc.) and often lead to the largest stellar masses, \lmass$>$10; 2) models with a flatter, grey attenuation law provide a better fit the UV region and lower stellar masses; 3) composite SEDs with a dusty galaxy and either a blue galaxy or a blue QSO dominating the SW bands also provide a good overall fit to SED and similar masses to 1). For scenarios where the LW bands are dominated by an obscured AGN, we find that models based on an obscured QSO plus a blue galaxy dominating the SW bands, or pure AGN models, where a combination of obscured and scattered emission by the torus dominates the whole SED, provide a good fit to the overall SED and lead to stellar masses for the galaxy host that are two orders of magnitude lower than in the dusty galaxy dominated scenarios, \lmass$=7-8$.
  
\item Without photometry at mid-IR wavelengths, SED modeling does not favor either of the galaxy-dominated or AGN-dominated scenarios. The unresolved, point-like sizes, on the other hand, are more suggestive of the AGN-dominated scenarios and disfavor those where a composite SED is caused by different stellar populations in distinct regions of a galaxy.

\item The number density arguments are also inconclusive since both scenarios might have some issues if any of the extreme properties are confirmed. The dusty galaxy scenario would imply an increase in the number density of very massive galaxies, \lmass$>10.5$, at z$>$7 of up to a factor of $\sim60$, relative to the pre-JWST estimates even if just a handful of them are confirmed to be that massive. For the AGN-dominated scenarios, the bolometric luminosities, derived by a simple dust correction of a blue QSO template, would imply an unexpectedly large number, 4$\times$10$^{-5}$ Mpc$^{-3}$, of obscured but intrinsically luminous QSOs, L$_{\rm bol}$=10$^{46-47}$ erg s$^{-1}$, at $z>7$, more than two orders of magnitude larger than current densities of bright, blue QSOs.   

\end{itemize}

\section*{Acknowledgments}

PGP-G acknowledges support  from  Spanish  Ministerio  de  Ciencia e Innovaci\'on MCIN/AEI/10.13039/501100011033 through grant PGC2018-093499-B-I00. \'AGA acknowledges the support of the Universidad Complutense de Madrid through the predoctoral grant CT17/17-CT18/17. This work has made use of the Rainbow Cosmological Surveys Database, which is operated by the Centro de Astrobiología (CAB), CSIC-INTA, partnered with the University of California Observatories at Santa Cruz (UCO/Lick, UCSC). This work is based on observations carried out under project number W20CK with the IRAM NOEMA Interferometer. IRAM is supported by INSU/CNRS (France), MPG (Germany) and IGN (Spain).

\software{Astropy \citep{astropy}, EAZY \citep{eazy}, GALFIT  \citep{galfit}, matplotlib \citep{matplotlib}, NumPy \citep{numpy}, PZETA (\citealt{pg08}), Prospector (\citealt{leja19} \citealt{johnson21}), Rainbow pipeline (\citealt{pg05,pg08}, \citealt{barro11b}), SExtractor \citep{sex}, Synthesizer (\citealt{pg05,pg08b})}

\bibliography{referencias}

\begin{appendix}
\section{Summary of modeling assumptions}
Here we provide additional details on the modeling assumptions adopted with each of the methods discussed in \S~\ref{s:stellarpop}.

{\bf For \texttt{EAZYpy}} \citep{eazy} we use the default template set  ``tweak fsps QSF 12 v3" which consists of a set 12 templates derived from the stellar population synthesis code FSPS \citep{conroy10}. The templates cover a wide range in age, dust attenuation, and log-normal star formation histories and they are computed using a \citet{chabrier} IMF and \citet{kriek13} dust attenuation law.

{\bf For \texttt{Synthesizer}} \citep{pg08b} we adopt the following assumptions: a delayed-exponential as the SFH, with timescale values $\tau$ between 100~Myr and 5 Gyr, ages between 1 Myr and the age of the Universe at the redshift of the source, all discrete metallicities provided by the \citet{bc03} models, a \citet{calzetti} attenuation law with V-band extinction values, A$_{\rm V}$ between 0 and 5 mag, and a \citet{chabrier} IMF. Nebular continuum and emission lines were added to the models as described in \citet{pg08b}. A Monte Carlo method was carried out in order to obtain uncertainties and account for degeneracies (see \citealt{dominguezsanchez16}). 

{\bf For \texttt{Prospector}} (\citealt{johnson21}; \citealt{leja19}) we adopt the following assumptions: we use the MIST stellar evolutionary tracks and isochrones \citep{choi16}, a \citet{chabrier} IMF, a range in stellar metallicity between -1.0 and 0.19, and gas phase metallicity between -2.0 and 0.5. The ionization parameter for the nebular emission ranges between -4. and -1.
The nebular line and continuum emission is generated using the \texttt{CLOUDY} \citet{cloudy} contained in FSPS and described in \citet{byler17}. 
For the attenuation law, we use either \citet{calzetti} or the more complex dust attenuation model which combines the \citet{cf00} two-component, birth-cloud vs. diffuse dust screens, approach with the \citet{kriek13} method that parametrizes the diffuse component as a combination of a Calzetti attenuation plus a Lorentzian Drude to model the strength of the UV bump. Both components are then modulated by a power-law factor that flattens or steepens the slope of the attenuation relative to Calzetti. The parameters being fit in this case are the ratio of the nebular to diffuse attenuation, which ranges between 0 and 2, but follows a clipped normal prior centered in 1,  and the dust index of the power law, which ranges between -1 and 0.4. One of the main advantages of Prospector is the possibility of using flexible star formation histories (SFHs). For the purpose of our analysis, we explore 2 options: 1) a fiducial delayed $\tau$-model, and a non-parametric piece-wise SFH. The 1-population $\tau$-model uses relatively wide priors on the stellar age, ranging between 1 Myr and the age of the Universe at the redshift of each source, and the star formation scale factor $\tau$ from 100 Myr to 20 Gyr. For the non-parametric model, we adopt the flexible SFH prescription \citep{leja19} with 6 time bins and the bursty-continuity prior \citep{tacchella22a}. This model includes 5 free parameters which control the ratio of SFR in six adjacent time bins; the first two bins are spaced at $0 - 5 $ Myr and $5 - 10$ Myr of lookback time, and the remaining four bins are log-spaced to a maximum age of 100 Myr.

{\bf For the hybrid galaxy - QSO model} we combine a galaxy component derived from \texttt{Prospector} using the fiducial $\tau$-model with the QSO templates from \citet{polletta06}. However, to provide more flexibility to the AGN component we include an additional degree of freedom in the \texttt{Prospector} modeling that includes AGN emission lines following the line ratios described in \citet{richardson14}.

\section{Table with the remaining 29 EROs}

\begin{figure*}
\centering
\includegraphics[width=16cm,angle=0]{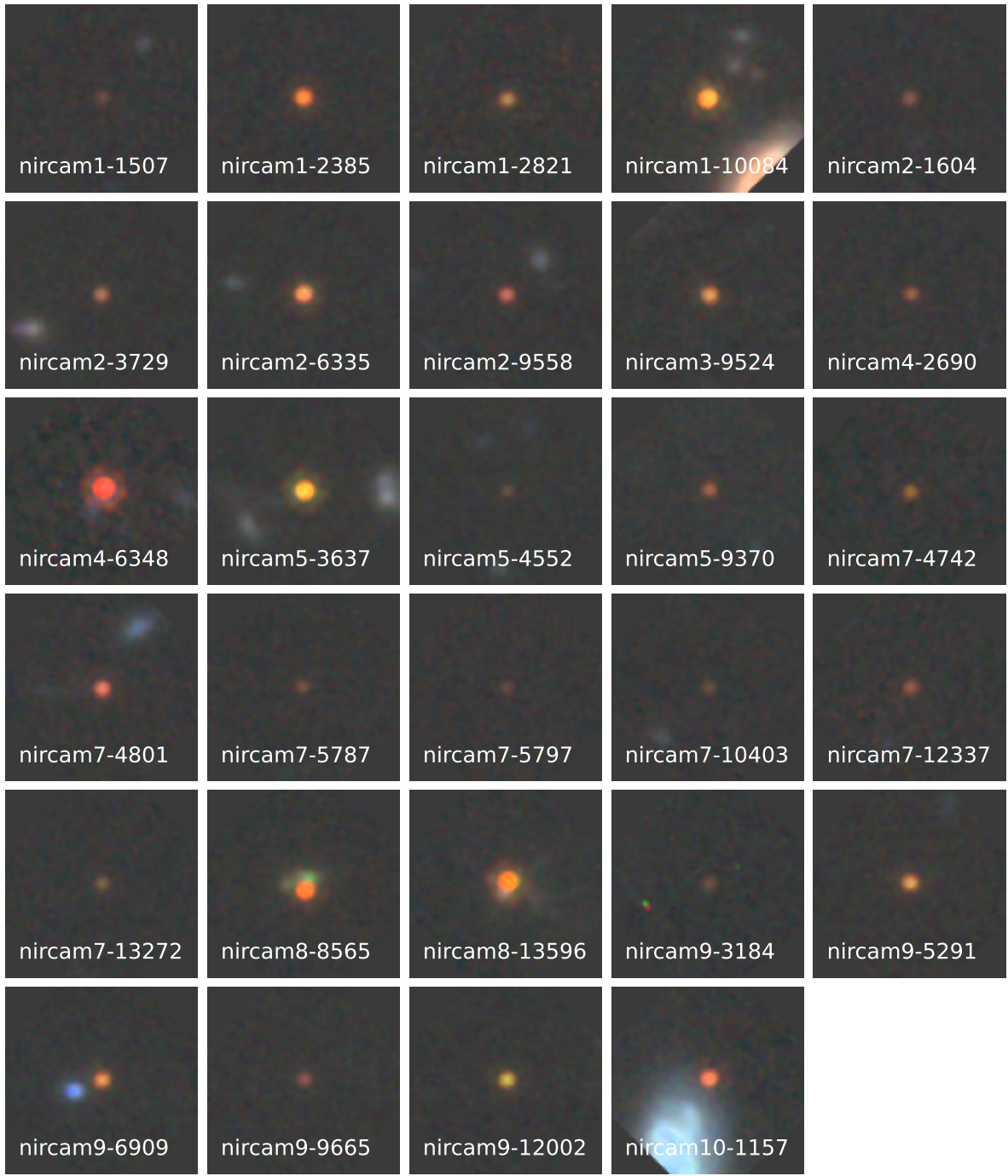}
\caption{Color composite (F277W+F356W+F444W) 2.5\arcsec$\times$2.5\arcsec cutouts of the 29 other EROs in the color selected sample. Similar to the 8 primary sources in Figures~\ref{fig:seds2} and \ref{fig:seds3}. These objects are also very red and remarkably homogeneous and compact.}
\label{fig:allsources}
\end{figure*}

Because of the interest in the peculiar sources for further analysis or follow-up, we list in Table~\ref{tab:otherobjects} the properties of the 29 remaining EROs in the color-selected sample of 37 objects, and we show 2.5\arcsec$\times$2.5\arcsec color composite cutouts for all them in Figure~\ref{fig:allsources}. In addition to the coordinates, we include in the table the photometric redshifts estimates with \texttt{EAZY} and the stellar masses used in the number density estimates of \S~\ref{s:discussion}. As discussed in \S~\ref{s:masses}, the values obtained with \texttt{FAST}, \texttt{Prospector}-np and \texttt{Synthesizer} provide a representative range of the variations in the stellar masses from the largest to the smallest values. We defer a more detailed, individualized analysis of the 29 sources, discussing the AGN-dominated scenarios, to a future paper (Kocevski in prep.).

\begin{table*}
%\big
\caption{Table with the remaining 29 EROs}
\label{tab:otherobjects}
\begin{tabular}{lcccccc}
\hline\hline
 ID  &  R.A.  &  Dec.  &  $z_{\rm phot}$  &  $\log$M$_{\star}$  &  $\log$M$_{\star}$ & $\log$M$_{\star}$  \\
   &   (deg)  &  (deg)  &     &  [M$_{\odot}$]  &  [M$_{\odot}$]  &  [M$_{\odot}$] \\
(1)  &  (2)  &  (3)  &  (4)  &  (5)  &  (6)  &  (7)  \\
\hline
nircam1-1507                  &  214.9372049 &   52.9653511 &  9.01$^{+ 0.27}_{- 1.53}$  &  10.23$\pm$0.20 &  10.29$^{+0.23}_{-0.21}$ &   9.71$\pm$0.21 \\
nircam1-2385$^{\star}$         &  214.9984072 &   53.0046186 &  6.49$^{+ 0.06}_{- 0.09}$  &  11.16$\pm$0.20 &  10.90$^{+0.06}_{-0.06}$ &   9.63$\pm$0.10 \\
nircam1-2821$^{\dagger,\star}$  &  214.9568340 &   52.9731536 &  5.86$^{+ 0.51}_{- 0.03}$  &   9.63$\pm$0.20 &   9.22$^{+0.51}_{-0.20}$ &   8.87$\pm$0.07 \\
nircam1-10084$^{\dagger\star }$ &  214.9830364 &   52.9560063 &  7.51$^{+ 0.75}_{- 0.00}$  &  11.10$\pm$0.20 &  10.81$^{+0.01}_{-0.02}$ &  10.76$\pm$0.07 \\
nircam2-1604$^{\dagger}$       &  214.9022374 &   52.9393697 &  8.62$^{+ 0.30}_{- 0.21}$  &   9.96$\pm$0.20 &   8.92$^{+0.18}_{-0.17}$ &   9.40$\pm$0.11 \\
nircam2-3729$^{\star}$         &  214.9257607 &   52.9456616 &  5.26$^{+ 0.09}_{- 0.09}$  &  10.02$\pm$0.20 &   9.87$^{+0.10}_{-0.09}$ &   9.22$\pm$0.09 \\
nircam2-6335$^{\star}$         &  214.9272433 &   52.9338926 &  5.89$^{+ 0.18}_{- 0.12}$  &  10.23$\pm$0.20 &  10.09$^{+0.05}_{-0.04}$ &   9.52$\pm$0.05 \\
nircam2-9558$^{\dagger}$        &  214.8761458 &   52.8808258 &  8.95$^{+ 0.15}_{- 0.24}$  &  10.53$\pm$0.20 &   9.74$^{+0.05}_{-0.05}$ &   9.93$\pm$0.06 \\
nircam3-9524$^{\dagger}$        &  214.8066661 &   52.8378071 &  6.49$^{+ 0.06}_{- 0.09}$  &   9.68$\pm$0.20 &   9.76$^{+0.02}_{-0.03}$ &   9.59$\pm$0.10 \\
nircam4-2690       &  214.6951501 &   52.7485639 &  8.62$^{+ 0.33}_{- 0.48}$  &  10.61$\pm$0.20 &  10.32$^{+0.16}_{-0.16}$ &   9.85$\pm$0.13 \\
nircam4-6348       &  214.7953672 &   52.7888465 &  5.20$^{+ 0.03}_{- 0.03}$  &  11.12$\pm$0.20 &  10.66$^{+0.09}_{-0.08}$ &  11.20$\pm$0.11 \\
nircam5-3637       &  214.8922437 &   52.8774066 &  7.24$^{+ 0.15}_{- 0.15}$  &  10.89$\pm$0.20 &  11.03$^{+0.02}_{-0.02}$ &  10.70$\pm$0.05 \\
nircam5-4552       &  214.8967641 &   52.8757973 &  6.10$^{+ 0.57}_{- 0.48}$  &   9.09$\pm$0.20 &   9.44$^{+0.50}_{-0.45}$ &   8.40$\pm$0.08 \\
nircam5-9370       &  214.9102805 &   52.8600731 &  7.21$^{+ 0.06}_{- 0.54}$  &   9.23$\pm$0.20 &   7.65$^{+0.03}_{-0.07}$ &   9.09$\pm$0.25 \\
nircam7-4742       &  215.1314689 &   52.9849141 &  7.48$^{+ 0.81}_{- 0.21}$  &  10.63$\pm$0.20 &  10.21$^{+0.10}_{-0.07}$ &   9.51$\pm$0.05 \\
nircam7-4801       &  215.1370670 &   52.9885588 &  5.20$^{+ 0.03}_{- 0.03}$  &  10.28$\pm$0.20 &   9.80$^{+0.10}_{-0.10}$ &   9.99$\pm$0.06 \\
nircam7-5787       &  215.0617802 &   52.9311768 &  7.39$^{+ 0.12}_{- 0.15}$  &  10.47$\pm$0.20 &   9.16$^{+0.28}_{-0.34}$ &   9.02$\pm$0.12 \\
nircam7-5797       &  215.1437160 &   52.9888945 &  8.32$^{+ 0.84}_{- 0.30}$  &   9.84$\pm$0.20 &   8.72$^{+0.25}_{-0.34}$ &   8.63$\pm$0.13 \\
nircam7-10403      &  215.1072079 &   52.9428524 &  6.34$^{+ 1.29}_{- 0.54}$  &   8.60$\pm$0.20 &   9.94$^{+0.33}_{-0.23}$ &   8.64$\pm$0.09 \\
nircam7-12337      &  215.1370160 &   52.9556504 &  6.52$^{+ 1.35}_{- 0.39}$  &   9.82$\pm$0.20 &   9.58$^{+0.22}_{-0.31}$ &   9.04$\pm$0.15 \\
nircam7-13272      &  215.0817101 &   52.9122515 &  5.68$^{+ 2.54}_{- 0.99}$  &  10.32$\pm$0.20 &  11.01$^{+0.38}_{-0.18}$ &   8.90$\pm$0.08 \\
nircam8-8565       &  214.9799601 &   52.8610789 &  6.40$^{+ 0.09}_{- 0.09}$  &  10.40$\pm$0.20 &  10.29$^{+0.02}_{-0.02}$ &   9.68$\pm$0.05 \\
nircam8-13596      &  215.0802967 &   52.9079028 &  7.48$^{+ 0.03}_{- 0.12}$  &  11.30$\pm$0.20 &  10.60$^{+0.01}_{-0.01}$ &  10.80$\pm$0.11 \\
nircam9-3184       &  214.8864225 &   52.8233786 &  6.55$^{+ 0.75}_{- 0.45}$  &   9.27$\pm$0.20 &   9.49$^{+0.48}_{-0.41}$ &   8.42$\pm$0.15 \\
nircam9-5291       &  214.8760316 &   52.8061093 &  5.77$^{+ 0.36}_{- 0.03}$  &   9.98$\pm$0.20 &   9.74$^{+0.12}_{-0.19}$ &   9.19$\pm$0.05 \\
nircam9-6909       &  214.8945555 &   52.8121629 &  5.71$^{+ 0.48}_{- 0.57}$  &  10.00$\pm$0.20 &  10.80$^{+0.06}_{-0.05}$ &  10.07$\pm$0.06 \\
nircam9-9665       &  214.8998043 &   52.8015414 &  5.17$^{+ 0.06}_{- 1.74}$  &  10.33$\pm$0.20 &   9.82$^{+0.08}_{-0.09}$ &   8.62$\pm$0.12 \\
nircam9-12002      &  214.8964700 &   52.7876884 &  7.30$^{+ 0.18}_{- 0.27}$  &  10.44$\pm$0.20 &   10.80$^{+0.07}_{-0.06}$ &  10.28$\pm$0.06 \\
nircam10-1157      &  214.8513502 &   52.7992928 &  7.21$^{+ 0.09}_{- 0.06}$  &   9.79$\pm$0.20 &   9.00$^{+0.70}_{-0.05}$ &   9.43$\pm$0.23 \\
\hline \hline
\end{tabular}
\tablecomments{
(1) Source ID in the CEERS catalog.
(2) Right ascension (J2000).
(3) Declination (J2000).
(4) Photometric redshift in \S~\ref{s:photoz}.
(5) Stellar masses derived using \texttt{FAST}.
(6) Stellar masses derived using \texttt{Prospector}-np.
(7) Stellar masses derived using \texttt{Synthesizer}.
$\dagger$ : This object is also studied in \citet{labbe23}.
$\star$ : This object is also studied in \citet{pgp23a}.
}
\end{table*}

\end{appendix}

\end{document}